\documentclass[%
 twocolumn,
superscriptaddress,
 amsmath,amssymb,
 aps,
16pt]{revtex4}

\usepackage{graphicx}
\usepackage{dcolumn}
\usepackage{bm}
\usepackage{braket}
\usepackage{comment}
\usepackage{mathbbol}
\usepackage{mathtools}
\usepackage{kbordermatrix}
\usepackage{color}

\newcommand*{\mydprime}{^{\prime\prime}\mkern-1.2mu}

\begin{document}

\preprint{APS/123-QED}

\title{Toward Witnessing Molecular Exciton Entanglement from Spectroscopy}

\author{Andrew E. Sifain}
\affiliation{%
 Department of Chemistry, Princeton University, Princeton, NJ, 08540, USA\\
}%
\author{Francesca Fassioli}
\affiliation{%
 Department of Chemistry, Princeton University, Princeton, NJ, 08540, USA\\
}%
\affiliation{%
 SISSA - Scoula Internazionale Superiore di Studi Avanzati, Trieste 34136, Italy
}%
\author{Gregory D. Scholes}%
 \email{gscholes@princeton.edu}
\affiliation{%
 Department of Chemistry, Princeton University, Princeton, NJ, 08540, USA\\
}%
\date{\today}

\begin{abstract}
Entanglement is a defining feature of quantum mechanics that can be a resource in engineered and natural systems, but measuring entanglement in experiment remains elusive especially for large chemical systems.  Most practical approaches require determining and measuring a suitable entanglement witness which provides some level of information about the entanglement structure of the probed state.  A fundamental quantity of quantum metrology is the quantum Fisher information (QFI) which is a rigorous witness of multipartite entanglement that can be evaluated from linear response functions for certain states.  In this work, we explore measuring the QFI of molecular exciton states of the first-excitation subspace from spectroscopy.  In particular, we utilize the fact that the linear response of a pure state subject to a weak electric field over all possible driving frequencies encodes the variance of the collective dipole moment in the probed state, which is a valid measure for QFI.  The systems that are investigated include the molecular dimer, $N$-site linear aggregate with nearest-neighbor coupling, and $N$-site circular aggregate, all modeled as a collection of interacting qubits.  Our theoretical analysis shows that the variance of the collective dipole moment in the brightest dipole-allowed eigenstate is the maximum QFI.  The optical response of a thermally equilibrated state in the first-excitation subspace is also a valid QFI.  Theoretical predictions of the measured QFI for realistic linear dye aggregates as a function of temperature and energetic disorder due to static variations of the host matrix show that 2- to 3-partite entanglement is realizable.  This work lays some groundwork and inspires measurement of multipartite entanglement of molecular excitons with ultrafast pump-probe experiments.
\end{abstract}

\maketitle


\section{Introduction}

Erwin Schr\"{o}dinger famously stated that entanglement is the characteristic trait of quantum mechanics with no classical analogue.\cite{schrodinger1935discussion}  An entangled quantum state of a composite many-body system may be completely known, whereas the states of its individual constituents are completely unknown--the state only exists as a whole.  Entanglement has been postulated since the birth of quantum mechanics, but considerable debate surrounded its validity on account of the Einstein-Podolsky-Rosen (EPR) paradox.\cite{einstein1935can}  However, the mistaken assumption of EPR--that physical reality is local--was later disproved with several experiments\cite{freedman1972experimental,aspect1981experimental,weihs1998violation} that violate John Bell's inequalities of local realism.\cite{bell1964einstein}  Entanglement has since been discovered to be a resource with potential to revolutionize next generation processes and technologies including quantum metrology,\cite{giovannetti2006quantum,giovannetti2011advances} teleportation,\cite{bouwmeester1997experimental,riebe2004deterministic,barrett2004deterministic} cryptography,\cite{gisin2002quantum} and computation.\cite{divincenzo1995quantum,nayak2008non,kassal2011simulating}  Entanglement also impacts quantum phase transitions,\cite{osterloh2002scaling,wu2004quantum,gu2004entanglement,dillenschneider2008quantum}  dynamics,\cite{bellomo2007non,nahum2017quantum} and energy conversion.\cite{thorwart2009enhanced,sarovar2010quantum,caruso2010entanglement,o2014non}  Despite its potential and prevalence in natural and engineered systems, entanglement's experimental realization remains elusive.

Two main factors contribute to the difficulty of measuring entanglement.  The first being that an entangled system is fragile and loses its quantumness as a result of interactions with its environment.  This process known as decoherence\cite{zurek2003decoherence} is the evolution of a coherent state into a statistical mixture.  As a result of these interactions, the state's ability to host entanglement is greatly suppressed. The second obstacle is that entanglement is quantified using information-theoretic measures such as purity or entropy.  These quantities are obtained by analyzing the state density matrix $\rho$, which is generally inaccessible from experiment.  Quantum state tomography--a method to find $\rho$ by a sequence of experiments--is only applicable to small systems made up of a handful of atoms.\cite{lanyon2017efficient} The goal from the viewpoint of chemistry is to design complex materials, from molecules to supramolecular assemblies, for quantum information science (QIS) applications.\cite{ferrando2016modular,wasielewski2020exploiting}  Thus, there is a need for practical approaches to measure entanglement, or more broadly, nonclassical correlations in large systems.\cite{ollivier2001quantum,fanchini2010non,modi2012classical}

Unlike measures that explicitly depend on $\rho$, a more practical approach for measuring entanglement in large systems is an entanglement \textit{witness}, where the evaluation of an operator in a given state may contain information about the state's entanglement structure.  The quantum Fisher information (QFI)\cite{liu2019quantum} is an entanglement witness and a fundamental quantity of quantum metrology.\cite{giovannetti2006quantum,toth2014quantum}  The QFI quantifies how much a state changes due to the interaction between an external agent and internal operator of the system. This \textit{distance} between the initial and final states as a result of the interaction relates to the maximum precision with which the (phase) parameter generated by the interaction can be estimated; the greater the sensitivity of the probed state to the generator, the greater precision in the estimated parameter.  As it turns out, the precision in the estimated parameter increases with the entanglement of the probed state.  Besides its use in parameter estimation, the QFI is a rigorous witness of multipartite entanglement\cite{pezze2009entanglement,toth2012multipartite,hyllus2012fisher} and provides signatures of quantum phase transitions.\cite{wang2014quantum,pezze2017multipartite} 

Inspired by the proposed mechanism of entanglement generation and evolution in photosynthetic light harvesting systems,\cite{olaya2008efficiency,fassioli2010distribution,sarovar2010quantum,caruso2010entanglement}  the focus of this work is on proposing a way to measure entanglement in molecular aggregates.  Molecular aggregates are assemblies of molecules with strong near-field Coulomb interactions between electronic excitations of individual molecules.  These interactions form exciton states--electronically excited states delocalized over spatially separated molecules.  Understanding the dynamics of molecular excitons gained significant attention\cite{fassioli2014photosynthetic, jumper2018coherent}  following the experimental reporting of wave-like beating in spectroscopic signals which showed the existence of long-lived coherences in photosynthetic complexes.\cite{engel2007evidence}  These experiments posed fundamental questions about whether the observed coherences are electronic in origin\cite{yuen2012witness,plenio2013origin,johnson2014practical,lim2019multicolor} and whether they play a functional role in the remarkably efficient energy transfer process of photosynthesis.\cite{duan2017nature}  While some issues may be resolved, such as how the interaction between the system and environment is not avoided to retain electronic coherences but rather exploited via a noise-assisted energy transfer process,\cite{plenio2008dephasing,cao2020quantum} photosynthetic light harvesting has undoubtedly opened interesting questions about possible links between coherence, nonclassical correlations, and functionality.\cite{scholes2017using,streltsov2017colloquium}  Here, we ask, can we measure some form of nonclassical correlation such as entanglement of a molecular exciton state?  

There exists important theoretical works that have addressed the topic of quantum state tomography of molecular excitons.  Namely, Yuen-Zhou and coworkers presented a method for characterizing the dynamics of a molecular dimer in the first-excitation subspace through a series of two-color photon-echo experiments.\cite{yuen2011quantum} Hoyer and coworkers extracted the time evolution of the excited state density matrix from the measured response in nonlinear pump-probe spectroscopy using a combined experimental and theoretical approach.\cite{hoyer2013inverting}  These works are ambitious and forward-thinking as they ultimately sought a full characterization of the quantum dynamics--more formally referred to as quantum process tomography in the QIS community. Yet there are setbacks in these works that open opportunities for future work such as extending the theory of Ref.~\cite{yuen2011quantum} to larger systems beyond the dimer and eliminating the use of a predetermined model in Ref.~\cite{hoyer2013inverting} describing how the system evolves following the probe interaction.

This work presents a theory based on the QFI to measure entanglement of molecular excitons from spectroscopy in the linear response.  We show that the interaction of the probing field and collective dipole moment encodes a QFI, and thus information about the probed state's entanglement structure.  We apply the theory to pure states as well as thermally equilibrated mixed states in the first-excitation subspace, the latter of which are experimentally feasible to probe.  We present results of realistic linear dye aggregates showing the dependence of the witnessed multipartite entanglement with temperature and energetic disorder, ranging from 2-partite and reaching 3-partite entanglement.  To our knowledge, this is the first proposal to measure, \textit{conclusively}, multipartite entanglement of excitons in nanoscale systems.

\section{Background}

\subsection{The Model}

Molecular aggregates are a collection of interacting chromophores commonly modeled by the Holstein Hamiltonian consisting of $N$ two-level sites each with an electronic ground and excited state.  Assuming neutral molecules, the chromophores are coupled through an electrostatic Coulomb interaction dominated by a dipole-dipole term.  The effects of the environment on the electronic system, \textit{i.e.}, electron-phonon interactions, are commonly modeled by coupling each site to a collection of local phonon modes.  The combined Hamiltonian ($\hbar=1$) is given by
\begin{multline}\label{eq:electron_phonon_hamiltonian}
H = \sum_{n}^{N}\omega_{n}\sigma_{n}^{+}\sigma_{n}^{-}+\sum_{m,n\ne m}^{N}J_{mn}\left(\sigma_{m}^{+}\sigma_{n}^{-}+\sigma_{m}^{-}\sigma_{n}^{+}\right)
\\
+\sum_{n}^{N}\sum_{k}\Omega_{k}a_{n,k}^{\dagger}a_{n,k}+\sum_{n}^{N}\sigma_{n}^{z}\sum_{k}g_{n,k}\left(a_{n,k}+a_{n,k}^{\dagger}\right)
\end{multline}
The first two terms make up the electronic system ($H_\text{sys}$) where $\sigma_{n}^{+}$ ($\sigma_{n}^{-}$) is the electronic raising (lowering) operator, $\omega_{n}$ is the electronic transition frequency, and $J_{mn}$ is the dipole-dipole coupling between sites $m\ne n$.  The third term of $H$ is the phononic contribution ($H_\text{env}$) where $a_{n,k}^{\dagger}$ ($a_{n,k}$) is the phononic creation (annihilation) operator associated to the $k$-th phonon mode coupled to site $n$ and $\Omega_{k}$ is the phononic frequency.  The last term of $H$ models linear electron-phonon coupling ($H_\text{sys-env}$) with coupling strength $g_{n,k}$, which is a function of the Huang-Rhys factor.\cite{mukamel1999principles}  The Hamiltonian of Eq.~\ref{eq:electron_phonon_hamiltonian} neglects interaction terms that couple site local ground and excited states since radiative and non-radiative decay to the excitonic ground state are assumed to occur on much longer timescales than other dynamics of interest.  Within this approximation, the number of electronic excitations is a conserved quantity $\left[\sum_{n}^{N}\sigma_{n}^{+}\sigma_{n}^{-}, H\right]=0$.  Therefore, the Hilbert space is a direct sum over different electronic excitation subspaces $H=H_{0}\oplus H_{1}\oplus H_{2}\oplus\cdots H_{N}$.  In this work, our inspiration comes from biomolecular systems where photoexcitation is either weak and/or doubly excited states are strongly suppressed.  Thus, we focus our attention on exciton dynamics confined to the one-excitation subspace $H_{1}$.

\subsection{Exciton States and Entanglement}

The most basic system illustrating delocalized and entangled exciton states in the absence of phononic degrees of freedom is a dimer ($N=2$).  The Hilbert space of chromophores $A$ and $B$ can be expanded in terms of site (localized) basis states $\ket{00}, \ket{10}, \ket{01}$, and $\ket{11}$ where $0$ and $1$ refer to zero and one excitations, respectively. For example, $\ket{01}$ hosts zero excitations on $A$ and one excitation on $B$.  The Hamiltonian of this purely electronic system is given by
\begin{equation}\label{h_dimer}
  H = \kbordermatrix{
    & \ket{00} & \ket{10} & \ket{01} & \ket{11} \\
    \ket{00} & 0 & 0 & 0 & 0 \\
    \ket{10} & 0 & \omega_{A} & -J & 0 \\
    \ket{01} & 0 & -J & \omega_{B} & 0 \\
    \ket{11} & 0 & 0 & 0 & \omega_{A}+\omega_{B}
  }
\end{equation}
The ground $\ket{\epsilon_{0}}=\ket{00}$ and doubly excited states $\ket{\epsilon_{3}}=\ket{11}$ are eigenstates with eigenenergies $0$ and $\omega_{A}+\omega_{B}$, respectively, whereas the coupling $J_{AB}=-J$ couples the singly excited states $\ket{10}$ and $\ket{01}$ giving rise to delocalized eigenstates
\begin{equation}\label{eq:exciton_states_dimer}
\begin{pmatrix} 
\ket{\epsilon_{1}} \\
\ket{\epsilon_{2}} 
\end{pmatrix}
=
\begin{pmatrix} 
\cos\theta & \sin\theta \\
-\sin\theta & \cos\theta 
\end{pmatrix}
\begin{pmatrix} 
\ket{10} \\
\ket{01} 
\end{pmatrix}
\end{equation}
with eigenenergies $\epsilon_{1,2}=\left(\omega_{A}+\omega_{B}\right)/2\mp\frac{1}{2}\sqrt{4J^{2}+\left(\omega_{A}-\omega_{B}\right)^{2}}$ and where $\theta$ is a mixing angle defined by $\tan 2\theta=2J/\left(\omega_{B}-\omega_{A}\right)$.  The degree of entanglement of the exciton states (Eq.~\ref{eq:exciton_states_dimer}) depends on $\theta$ which can be characterized with information-theoretic measures.  

One such measure is the quantum purity defined as $\mathrm{Tr}\left(\rho^{2}\right)$ which classifies pure and mixed states.  It is bounded by $1/d\le\mathrm{Tr}\left(\rho^{2}\right)\le 1$, where $d$ is the dimension of the Hilbert space with $\mathrm{Tr}\left(\rho^{2}\right)=1$ for pure states and $\mathrm{Tr}\left(\rho^{2}\right)<1$ for mixed states.  From a quantum information standpoint, all is known for pure states, whereas for mixed states, there is \textit{loss of information}.  The magnitude of $\mathrm{Tr}\left(\rho_{AB}^{2}\right)$ compared to that of one of its subsystems $\mathrm{Tr}\left(\rho_{A}^{2}\right)$ can determine whether partitions $A$ and $B$ are entangled by the conditions:
\begin{subequations}\label{eq:entanglement_condition_purity}
\begin{equation}
\mathrm{Tr}\left(\rho_{A}^{2}\right)<\mathrm{Tr}\left(\rho_{AB}^{2}\right)  
\end{equation}
\begin{equation}
\mathrm{Tr}\left(\rho_{B}^{2}\right)<\mathrm{Tr}\left(\rho_{AB}^{2}\right)  
\end{equation}
\end{subequations}
The loss of information in $\rho_{A}$ ($\rho_{B}$) with respect to $\rho_{AB}$ in Eq.~\ref{eq:entanglement_condition_purity} signifies that there exists quantum correlations between $A$ and $B$, making it impossible to completely know the state of $A$ ($B$) independently.  A density matrix of a subsystem or a reduced density matrix is calculated by averaging over the information pertaining to all other subsystems $\rho_{A}=\mathrm{Tr}_{B}\left(\rho_{AB}\right)$.  For bipartite systems, $\mathrm{Tr}\left(\rho_{A}^{2}\right)=\mathrm{Tr}\left(\rho_{B}^{2}\right)$.  For either of the exciton states in Eq.~\ref{eq:exciton_states_dimer}, 
\begin{equation}\label{eq:purity_exciton_states_dimer}
\text{Tr}\left(\rho_{A}^{2}\right)=\cos^{4}\theta+\sin^{4}\theta
\end{equation}
which is $<1$ unless $J=0$ ($\theta=\{0,\,\pi/2\}$).  In the limit of maximal mixing $\cos\theta=\sin\theta=1/\sqrt{2}$, the states are the well-known maximally entangled Bell states.\cite{nielsen2001quantum} 

\begin{figure}[h]
    \centering
    \includegraphics[width=0.45\textwidth]{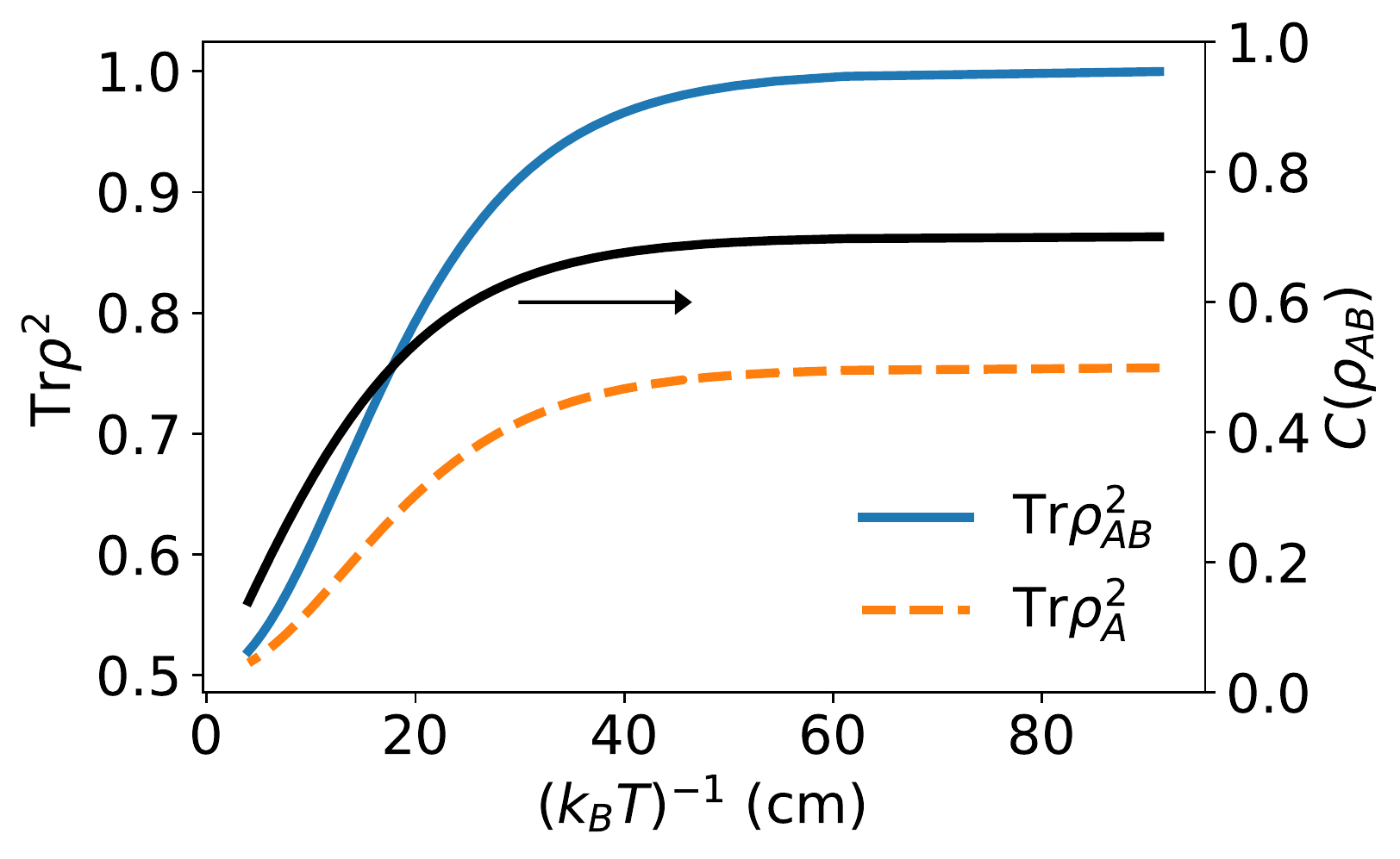}
    \caption{Entanglement of a dimer in a thermally equilibrated state in the first-excitation subspace as a function of inverse temperature.  Bipartite entanglement is computed with purity (left axis) and concurrence (right axis).  The state is entangled for all temperatures since $\mathrm{Tr}\rho_{A}^2<\mathrm{Tr}\rho_{AB}^2$ (Eq.~\ref{eq:entanglement_condition_purity}) and $C\left(\rho_{AB}\right)>0$ where the amount of entanglement increases with decreasing temperature.  Parameters of the dimer model are provided in the text.}
    \label{fig:dimer_thermal_exciton_purity_concurrence}
\end{figure}

Eq.~\ref{eq:electron_phonon_hamiltonian} models an electronic system coupled to a phononic bath at temperature $T$, which thermalizes the exciton.  In this work, we are interested in probing the entanglement of experimentally feasible (quasi) steady-states in the first-excitation subspace.  The excitonic density matrix in the long-time limit after averaging over phonons is $\rho_{AB}\left(t\rightarrow\infty\right)=e^{-\beta H_{1}}/\mathcal{Z}$ with inverse temperature $\beta=1/k_{B}T$ and partition function $\mathcal{Z}=\mathrm{Tr}\left(e^{-\beta H_{1}}\right)$.  To demonstrate the presence of entanglement and its dependence on temperature, we consider a dimer system whose parameters are inspired by the dominant dimeric structure in the Fenna-Matthews-Olson (FMO) complex given by sites 3 and 4.\cite{milder2010revisiting}  The site energies of the chromophores are $\omega_{A}=12328$ cm$^{-1}$ and $\omega_{B}=12472$ cm$^{-1}$ with dipole coupling $J=70.7$ cm$^{-1}$.  The purities calculated for the dimer and one of its subsystems are given by 
\begin{subequations}\label{eq:dimer_thermal_exciton_purity}
\begin{equation}
\text{Tr}\left(\rho_{AB}^{2}\right)=\frac{e^{-2\beta\epsilon_{1}}+e^{-2\beta\epsilon_{2}}}{\mathcal{Z}^{2}}
\end{equation}
\begin{multline}
\text{Tr}\left(\rho_{A}^{2}\right)=\frac{\left(e^{-\beta\epsilon_{1}}\cos^{2}\theta+e^{-\beta\epsilon_{2}}\sin^{2}\theta\right)^{2}}{\mathcal{Z}^{2}}\\
+\frac{\left(e^{-\beta\epsilon_{1}}\sin^{2}\theta+e^{-\beta\epsilon_{2}}\cos^{2}\theta\right)^{2}}{\mathcal{Z}^{2}}
\end{multline}
\end{subequations}
Eq.~\ref{eq:dimer_thermal_exciton_purity} is plotted in Fig.~\ref{fig:dimer_thermal_exciton_purity_concurrence} along with the concurrence $C\left(\rho_{AB}\right)$, commonly used to characterize the entanglement of two qubits in a mixed state and ranges from $C\left(\rho_{AB}\right)=0$ for a separable state to $C\left(\rho_{AB}\right)=1$ for a maximally entangled state.\cite{wootters1998entanglement}  Purity and concurrence show a similar relationship in which the state is not only entangled but the amount of entanglement increases with decreasing temperature, eventually leveling of at temperatures less than ${\sim}72$ K ($\beta={\sim}50$ cm).

But while entanglement measures such as purity and concurrence are insightful, they are difficult to measure experimentally because they explicitly depend on elements of $\rho$, which are not readily available.  Additionally, bipartite measures are unable to provide information about the multipartite entanglement structure of the state of interest with a single measurement.  For example, in order to establish that all subsystems of a state are entangled using a bipartite measure, all combinations of two partitions must be entangled.  This work brings attention to the QFI which is a rigorous witness of multipartite entanglement with relevance to electronic spectroscopy.  A key question that we pose and answer is the following: {\textit{Can spectroscopy be used to provide information about the entanglement of the probed state similar to that provided in Fig.~\ref{fig:dimer_thermal_exciton_purity_concurrence}?}}

\section{Theoretical Framework}
\subsection{Quantum Fisher Information} 
\label{quantum_isher_information}

The QFI is a fundamental quantity of quantum metrology.  It quantifies the maximal precision with which a parameter $\theta$ can be estimated in a given state.  Such parameters are generally explicit in the Hamiltonian and become encoded in the quantum state during the dynamics such as a driving magnetic field.\cite{pang2014quantum,pang2017optimal}  The variance of $\theta$ is bounded by the QFI through the quantum Cram\'{e}r-Rao bound, $\langle\left(\Delta \theta\right)^2\rangle \ge 1/\left(\nu F_{Q}\right)$, where $\nu$ is the number of independent measurements.  For a pure state $\ket{\psi}$,
\begin{equation}\label{qfi_pure_state_derivative}
F_{Q} = 4\left(\braket{\partial_{\theta}\psi|\partial_{\theta}\psi}+\left|\braket{\partial_{\theta}\psi|\psi}\right|^{2}\right)
\end{equation}
While $\theta$ can be more precisely estimated by increasing  $\nu$, repeated measurements on the same initially prepared state is difficult and often impractical.  Thus, increasing the QFI is a central task of quantum metrology.  The amount of $\theta$ information that is encoded in $\rho$ changes as the system evolves.  For infinitesimally small changes in $\theta$, the wavefunction can be represented by a unitary transformation $\ket{\psi_{\theta+\delta\theta}} = e^{-i\delta\theta\mathcal{O}}\ket{\psi_{\theta}}$, where $\mathcal{O}$ is the Hermitian generator of $\theta$.  That is, $\mathcal{O}$ is the observable that generates a unitary transformation of the state with respect to parameter $\theta$.  Eq.~\ref{qfi_pure_state_derivative} becomes
\begin{equation}\label{qfi_pure_state}
F_{Q} = 4\left(\bra{\psi}\mathcal{O}^{2}\ket{\psi}-\bra{\psi}\mathcal{O}\ket{\psi}^{2}\right)
\end{equation}
which is proportional to the variance of $\mathcal{O}$.  For a mixed state, the transformation of $\rho=\sum_{\lambda}p_{\lambda}\ket{\lambda}\bra{\lambda}$ is given by $\rho_{\theta+\delta\theta}=e^{-i\delta\theta\mathcal{O}}\rho_{\theta}e^{i\delta\theta\mathcal{O}}$ and the QFI takes a more complicated form
\begin{equation}\label{qfi_mixed_state}
F_{Q} = 2\sum_{\lambda, \lambda^{\prime}}\frac{\left(p_{\lambda}-p_{\lambda^{\prime}}\right)^{2}}{p_{\lambda}+p_{\lambda^{\prime}}}\left|\bra{\lambda}\mathcal{O}\ket{\lambda^{\prime}}\right|^{2}
\end{equation}
where the sum is over combinations of states $\ket{\lambda}$ and $\ket{\lambda^{\prime}}$ for which $p_{\lambda}+p_{\lambda^{\prime}} > 0$.  

The limit of the QFI differs in the case of separable versus entangled states.  Consider a system with $N$ identical sites and a local generator $\mathcal{O}=\sum_{i}^{N}\mathcal{O}_{i}$ where $\mathcal{O}_{i}$ acts on the $i$-th site.  The spectral width of $\mathcal{O}_{i}$ is $\left(\lambda_{M}-\lambda_{m}\right)$ where $\lambda_{M}$ and $\lambda_{m}$ are the largest and smallest eigenvalues of $\mathcal{O}_{i}$, respectively.  In the case of a separable state
\begin{equation}\label{qfi_limit_separable}
F_{Q}\le N\left(\lambda_{M}-\lambda_{m}\right)^{2}
\end{equation}
Saturating Eq.~\ref{qfi_limit_separable} is called the shot-noise limit $\langle\left(\Delta \theta\right)^2\rangle \ge 1/\left[\nu N\left(\lambda_{M}-\lambda_{m}\right)^{2}\right]$ whereas for an entangled state
\begin{equation}\label{qfi_limit_entangled}
F_{Q}\le N^{2}\left(\lambda_{M}-\lambda_{m}\right)^{2}
\end{equation}
Eq.~\ref{qfi_limit_entangled} shows that the QFI surpasses the shot noise limit $\langle\left(\Delta \theta\right)^2\rangle \ge 1/\left[\nu N^{2}\left(\lambda_{M}-\lambda_{m}\right)^{2}\right]$ and establishes the usefulness of entangled states for precision measurements.  Saturation of Eq.~\ref{qfi_limit_entangled} is called the Heisenberg limit and it can only be achieved with a maximally entangled state.  It is important to note that although a state may be entangled, detecting the state's entanglement depends on the choice of generator $\mathcal{O}$.  Thus, characterizing entanglement via QFI can be accomplished by maximizing the QFI, normalized by $\left(\lambda_{M}-\lambda_{m}\right)^{2}$, with respect to $\mathcal{O}$.   

The authors of Ref.~\cite{hyllus2012fisher} established bounds on the QFI for multipartite entanglement classes where the magnitude of QFI determines the number of entangled sites in a given state.  Fundamental to the proof is the concept of $n$-producibility.  A pure state is $n$-producible if it can be written as $\ket{\psi_\text{$n$-prod}}=\otimes_{l=1}^{M}\ket{\psi_{l}}$, where $\ket{\psi_{l}}$ is a state of $N_{l}\le n$ sites (such that $\sum_{l=1}^{M} N_{l}=N$).  A state is $n$-partite entangled if it is $n$-producible but not $\left(n-1\right)$-producible.  Therefore, a $n$-partite entangled state can be written as a product $\ket{\psi_\text{n-ent}}=\otimes_{l=1}^{M}\ket{\psi_{l}}$ containing at least one state with $N_{l}=n$ sites that does not factorize.  For example, the three body state $\ket{\psi_\text{2-ent}}=\ket{\phi}_{1}\otimes\ket{\chi}_{23}$ is two-partite entangled since $\ket{\chi}_{23}$ does not factorize.  The same condition holds for mixed states.

For a system with $N$ qubits, a general generator is given by
\begin{equation}\label{linear_two_mode_interferometer}
\mathcal{O}=\frac{1}{2}\sum_{i}^{N}\vec{n}_{i}\cdot\vec{\sigma}_{i}
\end{equation}
where $\vec{n}_{i}\cdot\vec{\sigma}_{i}$ acts on the $i$-th site. Here, $\vec{n}_{i}$ is a unit vector on the Bloch sphere and $\vec{\sigma}_{i}$ is a vector of the Pauli spin-1/2 matrices.  The following condition bounds $n$-producible states
\begin{equation}\label{multipartite_entanglement_condition}
F_{Q} \le sn^{2} + r^{2}
\end{equation}
where $s=\lfloor\frac{N}{n}\rfloor$ is the largest integer smaller than or equal to $\frac{N}{n}$ and $r=N-sn$.  A violation of Eq.~\ref{multipartite_entanglement_condition} proves $(n+1)$-partite entanglement.  This proof applies to spins greater than 1/2 provided that the spectra of local operators ($\lambda_{M}-\lambda_{m}$) are of unit width, as in the case of $\frac{1}{2}\left(\vec{n}_{i}\cdot\vec{\sigma}_{i}\right)$.

\subsection{QFI and Linear Response Theory}

Of central importance to this work is that the QFI is related to linear response theory and becomes a measurable quantity assuming that the chosen generator is experimentally accessible.  The QFI of a pure state can be expressed in terms of a response function.\cite{ozawa2019probing}  The same is true for thermal ensembles.\cite{hauke2016measuring}  Consider a material Hamiltonian, $H_{0}$, perturbed by a time-dependent interaction, $H\left(t\right)=H_{0}+f(t)\mathcal{O}$, where $\mathcal{O}$ is the interaction between the external agent and internal operator of the system and $f\left(t\right)$ is the time-dependence of the perturbation.  In the limit of weak coupling, 
the expectation value of $\mathcal{O}$ in the probed state $\rho$ is $\left<\mathcal{O}\left(t\right)\right>=\left<\mathcal{O}\right>+\int_{0}^{\infty}\mathrm{d}\tau f\left(t-\tau\right)R\left(\tau\right)$, where $R\left(\tau\right)$ is the linear response function.  The transformation of $R\left(\tau\right)$ to the frequency domain is
\begin{equation}
\chi\left(\omega\right)=\frac{i}{\hbar}\int_{0}^{\infty}\mathrm{d}t e^{i\omega t}\text{Tr}\left(\rho\left[\mathcal{O}\left(t\right),\mathcal{O}\left(0\right)\right]\right)
\end{equation}
where $\mathcal{O}\left(t\right)=e^{iH_{0}t/\hbar}\mathcal{O}e^{-iH_{0}t/\hbar}$.  Without loss of generality, we consider a state $\rho=\sum_{\lambda}p_{\lambda}\ket{\lambda}\bra{\lambda}$ where $\ket{\lambda}$ is an eigenstate of $H_{0}$ with energy $E_{\lambda}$.  The imaginary component of $\chi\left(\omega\right)$ is
\begin{equation}\label{chi_lehman_representation}
\chi{\mydprime}\left(\omega\right)=\pi\sum_{\lambda,\lambda^{\prime}}\left(p_{\lambda}-p_{\lambda^{\prime}}\right)\left|\bra{\lambda}\mathcal{O}\ket{\lambda^{\prime}}\right|^{2}\delta\left(\hbar\omega + E_{\lambda} -E_{\lambda^{\prime}}\right)
\end{equation}
The QFI of a pure state (Eq.~\ref{qfi_pure_state}) is an integral over $\chi{\mydprime}\left(\omega\right)$ with respect to driving frequencies $\omega$.  Assuming the system is in energy eigenstate $\ket{\psi}$ and $\bra{\psi}\mathcal{O}\ket{\psi}=0$, integrating Eq.~\ref{chi_lehman_representation} and taking the sum over $\lambda^{\prime}$ using the completeness relation $\sum_{\lambda^{\prime}}\ket{\lambda^{\prime}}\bra{\lambda^{\prime}}=\mathbb{1}$ gives
\begin{equation}\label{qfi_pure_state_in_terms_of_chi}
F_{Q} = \frac{4\hbar}{\pi}\int_{0}^{\infty}\mathrm{d}\omega\chi{\mydprime}\left(\omega\right)
\end{equation}
$\chi{\mydprime}\left(\omega\right)$ can also be expressed in terms of correlation functions, $\chi{\mydprime}\left(\omega\right)=\frac{1}{2\hbar}\left[\tilde{C}\left(\omega\right)-\tilde{C}\left(-\omega\right)\right]$ where $\tilde{C}\left(\omega\right)$ is the Fourier transform of $C\left(t\right)=\left<\mathcal{O}\left(t\right)\mathcal{O}\left(0\right)\right>$
\begin{equation}
\tilde{C}\left(\omega\right)=\int_{-\infty}^{\infty}\mathrm{d}t\left<\mathcal{O}\left(t\right)\mathcal{O}\left(0\right)\right>e^{i\omega t}
\end{equation} 
Due to dephasing and dissipation, the system decoheres and the correlation function $C\left(t\right)$ generally tends to zero.  A key feature that we will call attention to later in greater detail is that Eq.~\ref{qfi_pure_state_in_terms_of_chi} does not rely on any assumption about the form of dephasing and dissipation experienced by the system.

\subsection{QFI and Spectroscopy}
\label{fisher_information_and_spectroscopy}

The experimentally accessible generator from linear spectroscopy is the interaction between a weak (classical) electric field and the system's dipole moment $\mathcal{O}=-\vec{\mu}\cdot\vec{E}$ where $\vec{\mu}=\sum_{i}^{N}\vec{\mu_{i}}\sigma_{i}^{x}$ and $N$ is the number of sites.  We refer to $\mathcal{O}$ as the \textit{dipole-field generator}.  The spectrum of the dipole autocorrelation function is given by (with $\vec{E}=\hat{\mu}$)
\begin{equation}\label{spectrum_of_dipole_correlation}
I\left(\omega\right) = \int_{-\infty}^{\infty}\mathrm{d}t\left<\mu\left(t\right)\mu\left(0\right)\right>e^{i\omega t}
\end{equation}
which encodes transitions between the probed state $\ket{\psi}$ to all other states $\ket{\lambda}$ with nonzero transition dipole moment $\bra{\psi}\mu\ket{\lambda}\ne 0$ made up of contributions from stimulated emission and excited state absorption.  For pure probed states and $\bra{\psi}\mu\ket{\psi}=0$, the QFI can be expressed as
\begin{equation}\label{qfi_spectrum}
F_{Q}=\frac{2}{\pi}\int_{-\infty}^{\infty}\mathrm{d}\omega I\left(\omega\right)
\end{equation}
Eq.~\ref{qfi_spectrum} is a direct consequence of $\int_{-\infty}^{\infty}\mathrm{d}\omega e^{i\omega t}=2\pi\delta\left(t\right)$, where $\delta\left(t\right)$ is the Dirac delta function.  An alternative form of the QFI in terms of the symmetric correlation function, $S\left(\omega\right)=\frac{1}{2}\left(\tilde{C}\left(\omega\right)+\tilde{C}\left(-\omega\right)\right)$, is
\begin{equation}\label{qfi_symmetric_correlation}
F_{Q}=\frac{4}{\pi}\int_{0}^{\infty}\mathrm{d}\omega S\left(\omega\right)
\end{equation}

\subsection{QFI of Thermal Exciton State from Spectroscopy}

Although the QFI of a pure state can be evaluated from optical response, pure exciton states are very short lived as electronic energy converts to phonons (\textit{i.e.}, non-radiative decay) almost immediately following photoexcitation, precluding the measurement of their QFI. Here we show how the optical response of a quasi-equilibrium thermal vibronic state in the first-excitation subspace is also a valid QFI.  The system relaxes into a thermal state in the first-excitation subspace with reduced state $\rho_\text{sys}^\text{eq}=\sum_{n}^{N} p_{n}\ket{\epsilon_{n}}\bra{\epsilon_{n}}$ where 
\begin{equation}\label{eq:thermal_state}
p_{n}=
  \begin{cases}
      \exp\left(-\epsilon_{n}/k_{B}T\right)/\mathcal{Z} & \text{for $n\in \text{first-exc. subspace}$}\\
      0 & \text{otherwise}
    \end{cases} 
\end{equation}
and $\mathcal{Z}=\sum_{n}^{N}\exp\left(-\epsilon_{n}/k_{B}T\right)$.  Evaluating the QFI (Eq.~\ref{qfi_mixed_state}) of this state with the dipole-field generator results in
\begin{equation}
F_{Q}=4\sum_{m,n}^{N} p_{n}\left|\left<\epsilon_{n}|\mu|\epsilon_{m}\right>\right|^{2}
\end{equation}
where the sums over $m$ and $n$ run over all excitonic states including those outside of the first-excitation subspace.  The sum over $m$ is a completeness relation and the QFI becomes
\begin{equation}\label{qfi_thermal_exciton}
F_{Q}=4\sum_{n}^{N} p_{n}\left<\epsilon_{n}|\mu^{2}|\epsilon_{n}\right>
\end{equation}
which is precisely equal to the QFI determined from optical response (Eq.~\ref{qfi_spectrum}) with spectrum given by $I\left(\omega\right)=\int_{-\infty}^{\infty}\mathrm{d}t\,\text{Tr}\left[\rho_\text{sys}^\text{eq}\mu\left(t\right)\mu\left(0\right)\right]e^{i\omega t}$.  Importantly, Eq.~\ref{qfi_thermal_exciton} connects entanglement with the optical response of an experimentally realizable exciton state.  It is worth stressing that the QFI for a mixed state is complicated; Eq.~\ref{qfi_thermal_exciton} does not apply for general $\mathcal{O}$.  The reason the QFI is related to optical response for mixed states in the first-excitation subspace, such as $\rho_\text{sys}^\text{eq}$, is because the dipole-field generator connects states that differ by one excitation $\mathcal{O}\propto\sum_{i}^{N}\sigma_{i}^{x}$ where $\sigma_{i}^{x}=\sigma_{i}^{+}+\sigma_{i}^{-}$, and the QFI, which measures the dipole-allowed susceptibility, is strictly dependent on transitions between the populated first-excitation subspace and the unpopulated zero- and second-excitation subspaces.  In other words, the dipole-field generator does not induce transitions within the first-excitation subspace that is being probed.

\section{Results and Discussion}

\subsection{QFI of Dimer from Spectroscopy}
\label{sec:qfi_dimer_spectroscopy}

We now apply the theory to the dimer model given by Eq.~\ref{eq:electron_phonon_hamiltonian} with $N=2$.  The dipole-field generator is chosen to be $\mathcal{O}=\frac{1}{2}\sum_{i}^{N}\sigma_{i}^{x}$ which validates the $n$-partite entanglement relation of Eq.~\ref{multipartite_entanglement_condition}.  The generator expressed in terms of the exciton states is given by
\begin{equation}\label{transition_dipole_moment_initial}
\mathcal{O}=\alpha\ket{\epsilon_{0}}\bra{\epsilon_{1}}+\beta\ket{\epsilon_{0}}\bra{\epsilon_{2}}+\alpha\ket{\epsilon_{1}}\bra{\epsilon_{3}}+\beta\ket{\epsilon_{2}}\bra{\epsilon_{3}} + \text{H.c.}
\end{equation}
where $\alpha=\left(\cos\theta+\sin\theta\right)/2$ and $\beta=\left(\cos\theta-\sin\theta\right)/2$.  Eq.~\ref{transition_dipole_moment_initial} invokes the Franck-Condon approximation which states that phonons are slow compared to electrons during an electronic transition and therefore the dipole moment is a function of electronic degrees of freedom only.  The time dependence of $\mathcal{O}$ is determined with the unitary rotation $\mathcal{O}\left(t\right) = U^{\dagger}\mathcal{O}\left(0\right) U$ where $U$ is the time-evolution operator.  Expanding $U$ in the exciton basis,
\begin{equation}\label{time_evolution_operator}
U = \sum_{n}^{N} e^{-i\epsilon_{n}t/\hbar}\ket{\epsilon_{n}}e^{-iH_{n}^{\prime}t/\hbar}\bra{\epsilon_{n}}
\end{equation}
where elements of $U$ are eigenvalues with respect to excitons and operators with respect to phonons.  Without loss of generality, Eq.~\ref{time_evolution_operator} assumes sites are coupled to a distribution of global phonon modes such that bosonic operators $a_{n,k}=a_{k}$ and interaction strengths $g_{n,k}=g_{k}$ are independent of site index $n$.  In doing so, $U=e^{-iH_\text{sys}t/\hbar}e^{-i\left(H_\text{sys-env}+H_\text{env}\right)t/\hbar}$ since $\left[H_\text{sys}, H_\text{sys-env}+H_\text{env}\right]=0$.  $H_{n}^{\prime}$ is the phononic contribution in the $n$-th exciton state $e^{-i H_{n}^{\prime} t/\hbar}=\bra{\epsilon_{n}}e^{-i\left(H_\text{env} + H_\text{sys-env}\right)t/\hbar}\ket{\epsilon_{n}}$ which is diagonal in the exciton basis because the system operator $\sigma^{z}=\sum_{n}^N\sigma_{n}^{z}$ of the system-environment interaction shares a common eigenbasis with the electronic Hamiltonian $\left[\sigma^{z},\, H_\text{sys}\right]=0$.  Evaluating $\mathcal{O}\left(t\right)$,
\begin{align}
\begin{split}
\mathcal{O}\left(t\right) &= \alpha e^{i\left(\epsilon_{0}-\epsilon_{1}\right)t/
\hbar}\ket{\epsilon_{0}}e^{iH_{0}^{\prime}t/\hbar}e^{-iH_{1}^{\prime}t/\hbar}\bra{\epsilon_{1}}\\
&+ \beta e^{i\left(\epsilon_{0}-\epsilon_{2}\right)t/
\hbar}\ket{\epsilon_{0}}e^{iH_{0}^{\prime}t/\hbar}e^{-iH_{2}^{\prime}t/\hbar}\bra{\epsilon_{2}}\\
&+ \alpha e^{i\left(\epsilon_{1}-\epsilon_{3}\right)t/
\hbar}\ket{\epsilon_{1}}e^{iH_{1}^{\prime}t/\hbar}e^{-iH_{3}^{\prime}t/\hbar}\bra{\epsilon_{3}}\\
&+ \beta e^{i\left(\epsilon_{2}-\epsilon_{3}\right)t/
\hbar}\ket{\epsilon_{2}}e^{iH_{2}^{\prime}t/\hbar}e^{-iH_{3}^{\prime}t/\hbar}\bra{\epsilon_{3}}\\
&+ \text{H.c.}
\end{split}
\end{align}
The dipole autocorrelation function $C\left(t\right)=\left<\mu\left(t\right)\mu\left(0\right)\right>$ for the system initially prepared in $\ket{\epsilon_{0}}$ is given by
\begin{align}\label{eq:correlation_ground}
\begin{split}
C\left(t\right)&=\frac{\left(\cos\theta+\sin\theta\right)^{2}}{4} e^{i\left(\epsilon_{0}-\epsilon_{1}\right)t/\hbar}\langle e^{iH_{0}^{\prime}t/\hbar}e^{-iH_{1}^{\prime}t/\hbar}\rangle\\
&+ \frac{\left(\cos\theta-\sin\theta\right)^{2}}{4} e^{i\left(\epsilon_{0}-\epsilon_{2}\right)t/\hbar}\langle e^{iH_{0}^{\prime}t/\hbar}e^{-iH_{2}^{\prime}t/\hbar}\rangle
\end{split}
\end{align}
where the system and environment are assumed to be separable $\rho_\text{sys}\otimes\rho_\text{env}$ although this assumption is not a requirement for the theory to hold true, and $\langle\cdots\rangle$ denotes a thermal average over phonons.  The terms of Eq.~\ref{eq:correlation_ground} correspond to transitions from the ground state to the first-excitation subspace: $\ket{\epsilon_{0}}\rightarrow\ket{\epsilon_{1}}$ and $\ket{\epsilon_{0}}\rightarrow\ket{\epsilon_{2}}$.  The dephasing function $F_{mn}\left(t\right) = \langle e^{iH_{m}^{\prime}t/\hbar}e^{-iH_{n}^{\prime}t/\hbar}\rangle$ measures the overlap between the phonon wavepackets on $\ket{\epsilon_{m}}$ with the same phonon wavepackets initially prepared on $\ket{\epsilon_{n}}$.  The dephasing function modulates the energy gap $\epsilon_{n}-\epsilon_{m}$ and is the source of homogeneous broadening and vibronic progression.  $F_{mn}\left(t\right)$ is often approximated with the second-cumulant expansion
\begin{equation}\label{dephasing_function}
F_{mn}\left(t\right) = e^{-i\lambda_{mn} t/\hbar - g_{mn}\left(t\right)}
\end{equation}
with decaying part given by
\begin{multline}
g_{mn}\left(t\right) = \int_{0}^{\infty} \mathrm{d}\omega\rho_{mn}\left(\omega\right)\left[\coth\left(\hbar\omega/2k_{B}T\right)\left(\cos\omega t - 1\right)\right.
\\\left.+ i\sin\omega t\right]
\end{multline}
where $\lambda_{mn}=\hbar\int_{0}^{\infty} \mathrm{d}\omega \omega\rho_{mn}\left(\omega\right)$ is the reorganization energy and $\rho_{mn}\left(\omega\right)$ is the spectral density which encodes the distribution of normal mode frequencies weighted by the strength with which each mode couples to the $\ket{\epsilon_{m}}\rightarrow\ket{\epsilon_{n}}$ transition.\cite{mukamel1999principles}  Although $\rho_{mn}\left(\omega\right)$ significantly affects the lineshape, the QFI which is an integral over the spectrum given by Eq.~\ref{qfi_spectrum} is a conserved quantity on account of the Dirac delta function $\int_{-\infty}^{\infty}\mathrm{d}\omega e^{i\omega t}=2\pi\delta\left(t\right)$.  Thus, there are no limitations on the form of the probed pure state, the system-environment interaction $H_\text{sys-env}$, or $\rho_{mn}\left(\omega\right)$ to validate the theory.  The QFI of the ground state $\ket{\epsilon_{0}}$ is simply the sum of the prefactors in the correlation function (multiplied by $4$) resulting in $F_{Q}=2$, which is the maximum value for a separable state (\textit{i.e.}, shot-noise limit for $N=2$).  There is no choice of $\mathcal{O}$ that will give a QFI above this value.

\begin{figure}[h]
    \centering
    \includegraphics[width=0.45\textwidth]{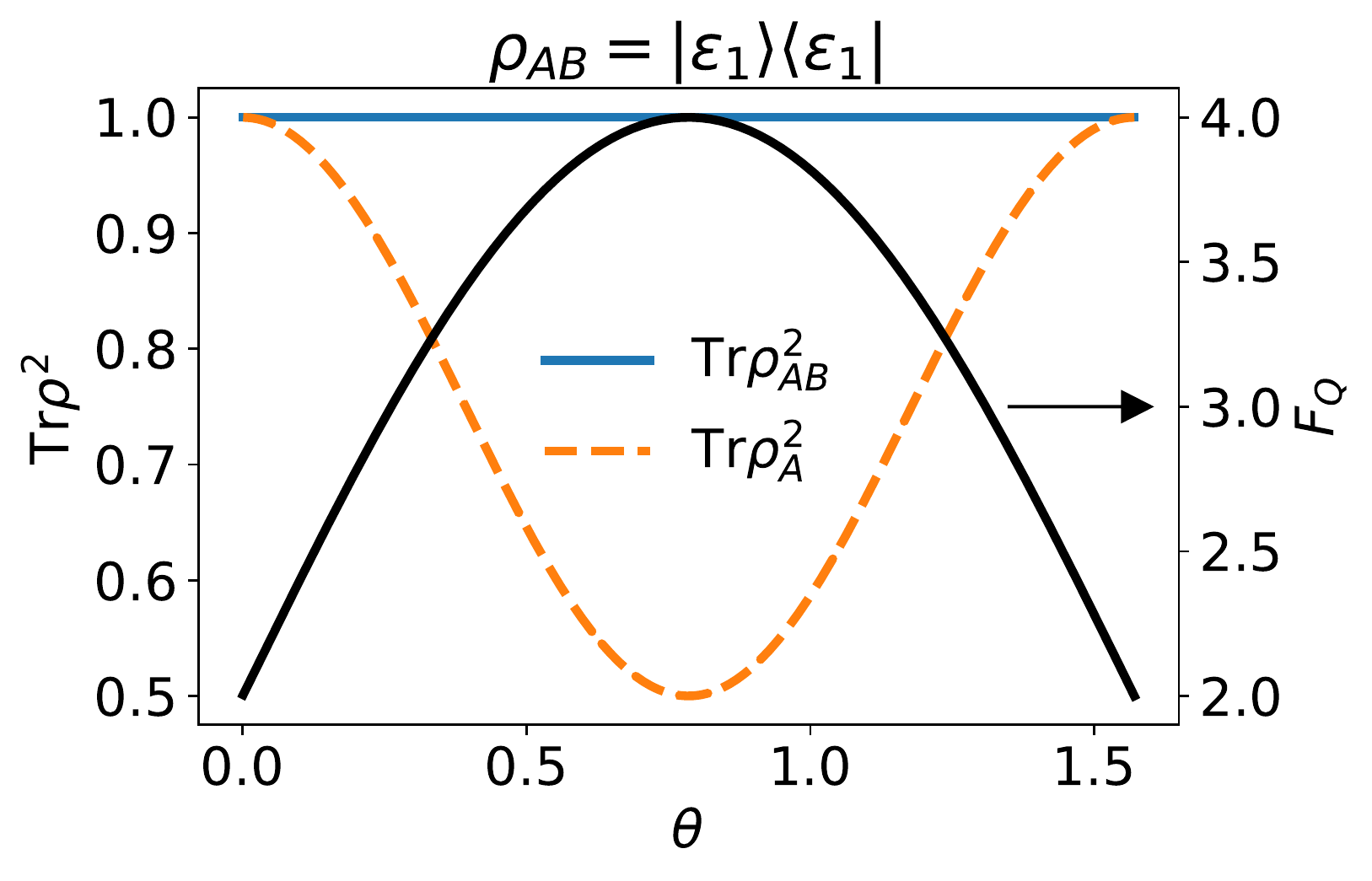}
    \caption{Entanglement measures of $\ket{\epsilon_{1}} = \cos\theta\ket{10} + \sin\theta\ket{01}$ evaluated with purity given by Eq.~\ref{eq:purity_exciton_states_dimer} (left axis) and QFI given by Eq.~\ref{eq:qfi_first_exciton} (right axis).  The generator used to compute the QFI is $\mathcal{O}=\frac{1}{2}\sum_{i}^{N}\sigma_{i}^{x}$.  $F_{Q}>2$ (for $N=2$) indicates the state is entangled.  The QFI (and purity) show maximum entanglement at the Bell state $\ket{\epsilon_{1}} = \frac{1}{\sqrt{2}}\ket{10} + \frac{1}{\sqrt{2}}\ket{01}$ and monotonically decrease (and increase) in the regime of large site energy mismatch $\left|\omega_{A}-\omega_{B}\right|\gg \left|J\right|$.} 
    \label{fig:entanglement_dimer_qfi}
\end{figure}

The QFI is more interesting in the case of $\ket{\epsilon_{1}}$, which unlike $\ket{\epsilon_{0}}$, is entangled.  The correlation function is
\begin{align}\label{eq:correlation_first_exciton}
\begin{split}
C\left(t\right)&=\frac{\left(\cos\theta+\sin\theta\right)^{2}}{4} e^{i\left(\epsilon_{1}-\epsilon_{0}\right)t/\hbar}\langle e^{iH_{1}^{\prime}t/\hbar}e^{-iH_{0}^{\prime}t/\hbar}\rangle\\
&+ \frac{\left(\cos\theta+\sin\theta\right)^{2}}{4} e^{i\left(\epsilon_{1}-\epsilon_{3}\right)t/\hbar}\langle e^{iH_{1}^{\prime}t/\hbar}e^{-iH_{3}^{\prime}t/\hbar}\rangle
\end{split}
\end{align}
The first term corresponds to stimulated emission $\ket{\epsilon_{0}}\leftarrow\ket{\epsilon_{1}}$ and the second term corresponds to excited-state absorption from the first- to second-excitation subspace $\ket{\epsilon_{1}}\rightarrow\ket{\epsilon_{3}}$.  The QFI is
\begin{align}\label{eq:qfi_first_exciton}
\begin{split}
F_{Q}&=2+4\sin\theta\cos\theta\\
&=2+\frac{4J}{\sqrt{4J^{2}+(\omega_{A}-\omega_{B})^{2}}}
\end{split}
\end{align}
Here, we see how the QFI depends on site energies and coupling.  The minimum QFI occurs if $J=0$ corresponding to a separable state.  The maximum QFI occurs if $\omega_{A}=\omega_{B}$, $J>0$ ($\theta=\pi/4$) corresponding to one of the famous Bell states.  Eq.~\ref{eq:qfi_first_exciton} and purity (Eq.~\ref{eq:purity_exciton_states_dimer}) are shown in Fig.~\ref{fig:entanglement_dimer_qfi} for comparison.

\begin{figure*}[t]
    \centering
    \includegraphics[width=0.95\textwidth]{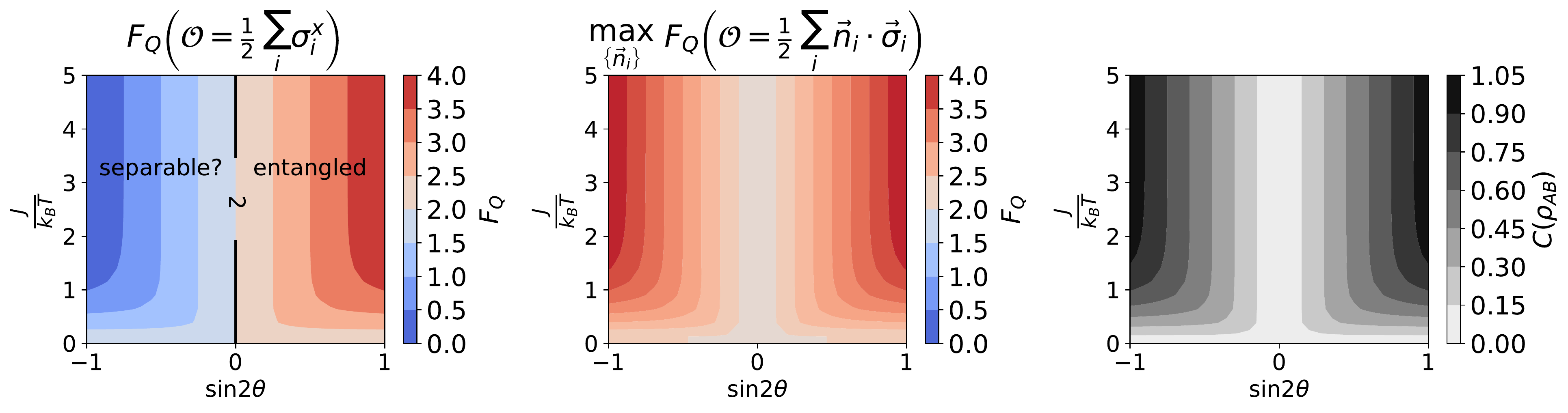}
    \caption{Heatmap of entanglement measures for a thermally equilibrated exciton state of the dimer ($N=2$) as a function of $\sin2\theta=2J/\sqrt{4J^{2}+\left(\omega_{A}-\omega_{B}\right)^{2}}$ and $J/k_{B}T$.  Entanglement measures include the QFI computed with the dipole-field generator (left panel), the maximum QFI (middle panel), and the concurrence (right panel).  The dipole-field generator is able to detect an entangled state for $J>0$ only, whereas the maximum QFI and concurrence  show that the state is entangled for all $J\ne 0$, and the degree of entanglement is symmetric around $J=0$.} 
    \label{fig:thermal_exciton_entanglement_dimer}
\end{figure*}

Although a given state may be entangled, its entanglement is not necessarily detected via a dipole-field interaction.  A worthwhile inquiry therefore is to evaluate the maximum QFI and determine how effective the dipole-field generator is at witnessing the entanglement.  The maximum QFI can be evaluated by taking the general two-level generator (Eq.~\ref{linear_two_mode_interferometer}) and maximizing the result with respect to site Bloch vectors $\vec{n}_{1}$ and $\vec{n}_{2}$.  The QFI in terms of $\vec{n}_{1}$ and $\vec{n}_{2}$ is
\begin{multline}\label{qfi_in_cartesian_vecs_dimer}
F_{Q} = 2 \pm 2\left(n_{1}^{x}n_{2}^{x} + n_{1}^{y}n_{2}^{y}\right)\sin2\theta - 2n_{1}^{z}n_{2}^{z} 
\\ - \left(n_{1}^{z}-n_{2}^{z}\right)^{2}\cos^{2}2\theta
\end{multline}
where the ``$+$" and ``$-$" solutions are those of the exciton states $\ket{\epsilon_{1}}$ and $\ket{\epsilon_{2}}$, respectively.  The $\hat{x}$ and $\hat{y}$ components must be parallel or antiparallel depending on the sign of $\pm\sin2\theta$.  By setting $n_{1}^{x}n_{2}^{x} + n_{1}^{y}n_{2}^{y}=\text{sgn}\left(\pm\sin2\theta\right)-n_{1}^{z}n_{2}^{z}$ and maximizing with respect to $n_{1}^{z}$ and $n_{2}^{z}$, one finds $n_{1}^{z}=n_{2}^{z}=0$.  The maximum QFI for both $\ket{\epsilon_{1}}$ and $\ket{\epsilon_{2}}$ is 
\begin{equation}
F_{Q}^{\text{max}}=2 + \frac{4\left|J\right|}{\sqrt{4J^{2} + \left(\omega_{A}-\omega_{B}\right)^{2}}}
\end{equation}
While the single-excitation states host the same degree of entanglement, the choice of $\vec{n}_{1}\cdot\vec{n}_{2}$ that maximizes the QFI depends on the sign of $J$.  The maximum QFI for $J>0$ is achieved for $\vec{n}_{1}\cdot\vec{n}_{2}=1$ for $\ket{\epsilon_{1}}$ and $\vec{n}_{1}\cdot\vec{n}_{2}=-1$ for $\ket{\epsilon_{2}}$.  This condition is inverted if $J<0$.  In summary, the single-excitation states are entangled if $J\ne 0$, but the measured QFI via the dipole-field interaction is the maximum QFI only in the case of $\ket{\epsilon_{1}}$ for $J>0$ and $\ket{\epsilon_{2}}$ for $J<0$.  To further demonstrate, the QFI of $\ket{\epsilon_{2}}$ computed with the dipole-field generator is
\begin{align}\label{eq:qfi_second_exciton}
\begin{split}
F_{Q}&=2-4\sin\theta\cos\theta\\
&=2-\frac{4J}{\sqrt{4J^{2}+(\omega_{A}-\omega_{B})^{2}}}
\end{split}
\end{align}
which shows that entanglement is not detected ($F_{Q}\le 2$) if $J\ge 0$ and detected ($F_{Q}>2$) if $J<0$.

But for all practical purposes, an initially prepared pure state will relax very quickly on an ultrafast timescale, precluding the state from being probed.  The system will relax non-radiatively into lower-lying exciton states until it reaches a thermal state in the first-excitation subspace (Eq.~\ref{eq:thermal_state}).  Remarkably, the optical response of this thermal state is still a valid QFI (Eq.~\ref{qfi_thermal_exciton}).  Evaluating Eq.~\ref{qfi_thermal_exciton} for the dimer with the dipole-field generator gives (Appendix~\ref{qfi_thermal_exciton_dimer})
\begin{equation}\label{eq:qfi_dimer_thermal_state}
F_{Q} = 2 + 2\tanh\left(\frac{1}{k_{B}T}\left|\frac{J}{\sin2\theta}\right|\right)\sin2\theta
\end{equation}
This result is plotted in Fig.~\ref{fig:thermal_exciton_entanglement_dimer} (left panel) in the form of a heatmap as a function of $\sin2\theta$ given in Eq.~\ref{eq:qfi_first_exciton}, which quantifies both the sign of the coupling and the mismatch in site energies, and the ratio of the coupling to thermal energy $J/k_{B}T$.  The QFI computed with the dipole-field generator is an entanglement witness only in the regime $J>0$ where $F_{Q}>2$.  For $J<0$ however, entanglement is not detected.  The maximum QFI (Appendix~\ref{qfi_thermal_exciton_dimer})
\begin{equation}\label{eq:max_qfi_thermal_exciton_dimer}
F_{Q}^\text{max} = 2 + 2\tanh\left(\frac{1}{k_{B}T}\left|\frac{J}{\sin2\theta}\right|\right)\left|\sin2\theta\right|
\end{equation}
shows that the degree of entanglement is symmetric around $J=0$ (Fig.~\ref{fig:thermal_exciton_entanglement_dimer} (middle panel)).   The left and middle panels of Fig.~\ref{fig:thermal_exciton_entanglement_dimer} differ because the low-energy state, which is more populated at thermal equilibrium, is the bright state when $J>0$; the optical response of this thermal state is dependent on the dipole susceptibility of this low-energy state.  On the other hand when $J<0$, the high-energy state dominates the response, but there is a competing effect in which increasing the high-energy state's population with increasing temperature also lessens the state's entanglement.  Although the response increases with temperature, it does not surpass the shot-noise limit and therefore the state's entanglement cannot be witnessed via the dipole-field interaction.  The concurrence shown in Fig.~\ref{fig:thermal_exciton_entanglement_dimer} (right panel) is in agreement with the maximum QFI in Fig.~\ref{fig:thermal_exciton_entanglement_dimer} (middle panel).

\subsection{QFI of Linear Aggregate with Nearest-Neighbor Coupling}
\label{linear_chain_circular_aggregate}

\begin{figure*}[t]
    \centering
    \includegraphics[width=0.95\textwidth]{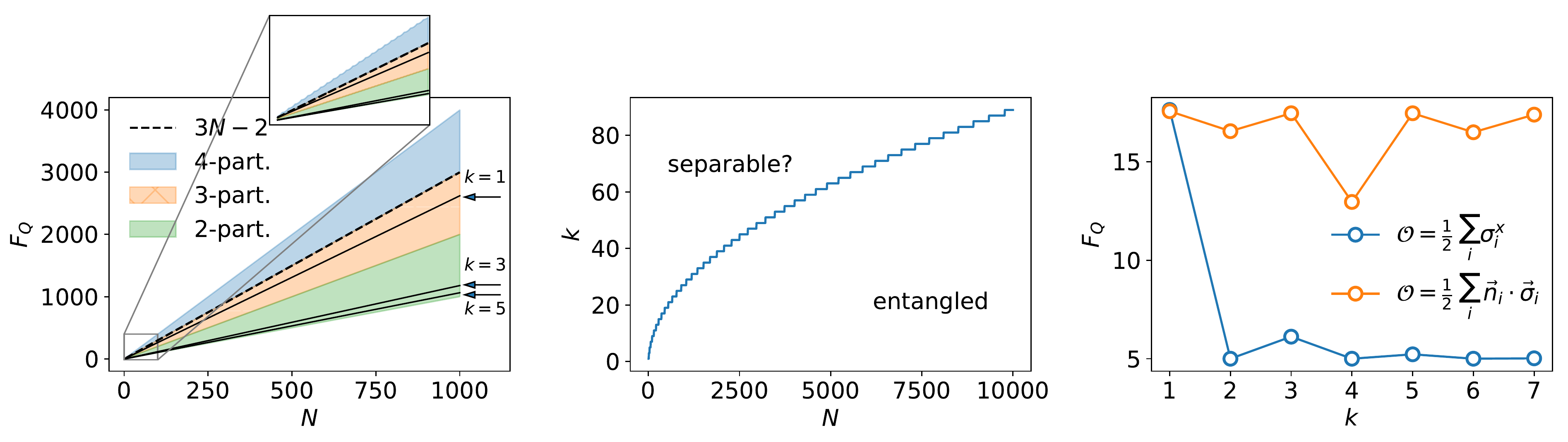}
    \caption{QFI of the linear aggregate with nearest-neighbor coupling.  (Left panel) QFI as a function of number of sites $N$ for the $k=1$, $3$, and $5$ states computed with the dipole-field generator.  The maximum QFI of the brightest state ($k=0$) of the circular aggregate given by $3N-2$ and the bounds of $n$-partite entanglement classes (Eq.~\ref{multipartite_entanglement_condition}) are also shown.  Only the $k=1$ state is in the regime of 3-partite entanglement, and it is bounded from above by $3N-2$ for all $N>2$.  The upper bound is reached only for the dimer ($N=2$). (Middle panel) The $k=k^{\prime}$ state that splits the band between states whose entanglement can and cannot be witnessed via a dipole-field interaction. (Right panel) A comparison between the QFI computed with the dipole-field generator and the numerically obtained maximum QFI (using Bayesian optimization) for an $N=7$ aggregate shows that the QFI of the brightest state ($k=1$) is maximized with the dipole-field generator.} 
    \label{fig:nn_periodic_aggregates_dipole_bright-state_n-partite}
\end{figure*}

We now focus our discussion on larger aggregates with arbitrary $N>2$.  The exciton states and energies of the pristine linear aggregate with identical site energies $\omega_{n}=\omega$ and nearest-neighbor couplings $J_{mn}=-J\delta_{m,n\pm 1}$ with open boundary conditions are
\begin{subequations}
\begin{equation}\label{eigenstates_linear_aggregate_nn_coupling}
\ket{\psi_{k}^{L}} = \sqrt{\frac{2}{N+1}}\sum_{n=1}^{N} \sin\left(\frac{\pi k n}{N+1}\right)\ket{n}
\end{equation}
\begin{equation}
\epsilon_{k}^{L}=-2J\cos\left(\frac{\pi k}{N+1}\right)
\end{equation}
\end{subequations}
for $k=1,\,2,...\,N$.\cite{fidder1991optical}  The QFI of state $k$ with the dipole-field generator is (Appendix~\ref{qfi_linear_aggregate_derivation})
\begin{equation}\label{eq:qfi_linear_aggregate_nn_coupling}
F_{Q}^{k}=\left(N-2\right)+2\left(\frac{1-\left(-1\right)^{k}}{N+1}\right)\frac{\sin^{2}\left(\frac{\pi k N}{2\left(N+1\right)}\right)}{\sin^{2}\left(\frac{\pi k}{2\left(N+1\right)}\right)}
\end{equation}
Fig.~\ref{fig:nn_periodic_aggregates_dipole_bright-state_n-partite} (left panel) shows Eq.~\ref{eq:qfi_linear_aggregate_nn_coupling} as a function of $N$ for $k=1$, $3$, and $5$ as well as the $n$-partite entanglement classes (Eq.~\ref{multipartite_entanglement_condition}).  The QFI is largest for $k=1$ corresponding to 3-partite entanglement.  The QFI decreases with increasing $k$ and, for a given $N$, there exists a $k=k^{\prime}$ above which the QFI falls below the shot-noise limit $F_{Q}=N$; the entanglement of states $k>k^{\prime}$ can no longer be detected with the dipole-field generator.  The smallest $k^{\prime}$ such that $F_{Q}^{k^{\prime}}>N$, distinguishing the entangled and separable regimes, is shown in Fig.~\ref{fig:nn_periodic_aggregates_dipole_bright-state_n-partite} (middle panel).

Unfortunately, we were unable to obtain an analytical solution of the maximum QFI for linear aggregates with general $N>2$.  Instead, we present results of a numerically optimized QFI in a small aggregate with $N=7$ shown in Fig.~\ref{fig:nn_periodic_aggregates_dipole_bright-state_n-partite} (right panel).  The maximum QFI of the $k=1$ state is obtained with the dipole-field generator, but for $k>1$, the QFI decreases exponentially (similar to Fig.~\ref{fig:nn_periodic_aggregates_dipole_bright-state_n-partite} (left panel)).  The maximum QFI is symmetric around the center of the band ($k=4$) with maximum entanglement at the band edges and minimum entanglement at the center; the dipole-field generator is limited to characterizing the $k=1$ state only.  Note the parallels between the QFI computed with the dipole-field generator and the ground-to-excited state transition dipole moment (squared) given by
\begin{equation}\label{eq:linear_aggregate_transition_dipole}
\left(\mu_{k}^{L}\right)^{2}=\mu^{2}\left(\frac{1-\left(-1\right)^{k}}{N+1}\right)\cot^{2}\left(\frac{\pi k}{2\left(N+1\right)}\right)
\end{equation}
which relates directly to stimulated emission.\cite{fidder1991optical}   The $k=1$ state is both strongly susceptible to the dipole-field generator and it is the brightest state of the system.  The susceptibility of a general state $k$ asymptotically scales with its transition dipole moment, and similar to $\left(\mu_{k}^{L}\right)^{2}$ which drops off as $k^{-2}$ for $k\ll N$, so too does Eq.~\ref{eq:qfi_linear_aggregate_nn_coupling}. 

To provide more rigor to the claim that the maximum QFI of the $k=1$ state can be obtained with the dipole-field generator, we turn to aggregates with periodic boundary conditions (\textit{i.e.}, circular aggregates) with indistinguishable sites and couplings that are not limited to nearest-neighbors.  The main motivation to study circular aggregates is that natural photosynthetic proteins such as LH1 and LH2 have a circular architecture.\cite{olaya2008efficiency}  Additionally, circular aggregates are often used to model linear aggregates with $N\gg1$ because of their simple exciton states given by
\begin{equation}\label{eq:eigenstates_circular_aggregate}
\ket{\psi_{k}^{C}}=\frac{1}{\sqrt{N}}\sum_{n=1}^{N}e^{2\pi i\left(kn/N\right)}\ket{n}
\end{equation}
for $k=0,\,1,...\,N-1$.\cite{fidder1991optical}  Unique to the circular aggregate is that oscillator strength is concentrated in the superradiant $k=0$ state, which is due to permutational invariance among the sites. This symmetry is broken in the linear aggregate because sites are not equivalent, and oscillator strength is distributed among a number of low $k$ states (Eq.~\ref{eq:linear_aggregate_transition_dipole}).  In the limit of large $N\gg 1$ , the circular and linear aggregates exhibit identical spectroscopic properties because the low $k$ bright states of the linear aggregate become degenerate with its $k=1$ state.  The maximum QFI of the $k=0$ state is as an upper bound on the maximum QFI for the $k=1$ state of the linear aggregate for all $N$.  To that end, we were able to find that the dipole-field generator maximizes the QFI of the $k=0$ state resulting in $F_{Q}=3N-2$ (Appendix~\ref{maximum_qfi_k=0_state_circular_aggregate}).  This result is plotted in Fig.~\ref{fig:nn_periodic_aggregates_dipole_bright-state_n-partite} (left panel) along with Eq.~\ref{eq:qfi_linear_aggregate_nn_coupling}, confirming that the $k=1$ state of the linear aggregate is limited to 3-partite entanglement.

Delocalization is a common measure in the context of molecular excitons which begs for consideration of how delocalization and multipartite entanglement compare to one another.  We believe that their correspondence is ambiguous or at least unknown.  Let us take for example our analysis of the $k=0$ state of Eq.~\ref{eq:eigenstates_circular_aggregate} which was shown in Appendix~\ref{maximum_qfi_k=0_state_circular_aggregate} to be 3-partite entangled for $N>3$.  Delocalization of a general state $\ket{\psi}=\sum_{i}^{N} c_{i}\ket{\psi_{i}}$, with expansion coefficients $c_{i}$ and basis states $\ket{\psi_{i}}$, can be quantified with the participation ratio (PR) expressed as $\left(\sum_{i}^{N}\left|c_{i}\right|^{2}\right)^{2}\big/\sum_{i}^{N}\left|c_{i}\right|^{4}$, ranging from 1 (localized to a single site) to $N$ (equally delocalized across all $N$ sites).\cite{scholes2019limits}  The PR of the $k=0$ state of Eq.~\ref{eq:eigenstates_circular_aggregate} is $N$ which clearly grows with $N$, whereas the multipartite entanglement does not.  While $\text{PR}>1$ signifies at least bipartite entanglement, beyond that, how PR relates to multipartite entanglement is not clear.  Therefore, entanglement is likely fundamentally different from delocalization.

\begin{figure}[h]
    \centering
    \includegraphics[width=0.45\textwidth]{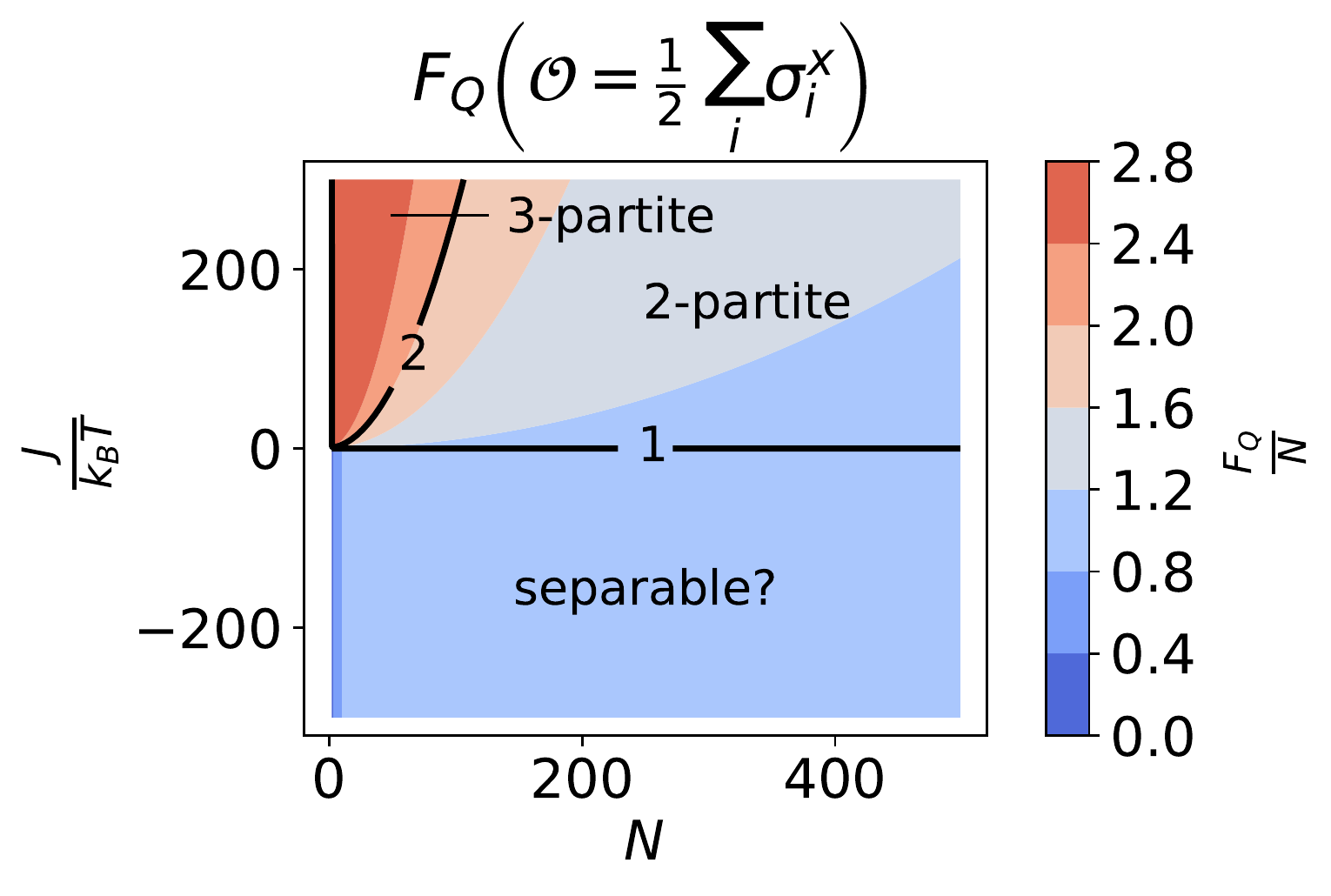}
    \caption{Heatmap of the QFI per number of sites $F_{Q}/N$ for a thermally equilibrated exciton state of the linear aggregate with nearest-neighbor coupling as a function of $N$ and $J/k_{B}T$.  The entanglement of aggregates with $J>0$ can be realized with the dipole-field generator, whereas the entanglement of aggregates with $J<0$ cannot.  The witnessed entanglement is at least 2-partite for all $N\ge 2$, and at most 3-partite.} 
    \label{fig:thermal_exciton_entanglement_nn}
\end{figure}

For a thermally equilibrated exciton state of the linear aggregate, we find that for $J<0$ the state's entanglement cannot be detected from linear spectroscopy (Fig.~\ref{fig:thermal_exciton_entanglement_nn}).  Again, this effect is because the brightest state is located at the top of the band which hosts very little population and only becomes populated with increasing temperature.  But increasing temperature is also detrimental to the entanglement of the state.  Therefore, although the optical response increases with temperature, the entanglement cannot be detected from linear spectroscopy even if it exists.  This scenario is unlike $J>0$ for which the brightest state is at the bottom of the band.  Witnessing 3-partite entanglement rapidly decreases with increasing $N$ because the low-energy bright states become closer in energy and their Boltzmann populations become more alike.  As a result, the relative population of the brightest $k=1$ state decreases and the contribution of this state to the overall optical response decreases as well.

\subsection{QFI of Linear Dye Aggregates with Static Energy Disorder}

We shall now model the measured QFI for realistic aggregates.  The systems under consideration are prototypical J aggregates ($J>0$) of the dye pseudoisocyanine (PIC) whose spectroscopic properties are understood from theory.  The authors of Ref.~\cite{heijs2005decoherence} modeled the temperature-dependent absorption linewidth of such aggregates and found excellent quantitative agreement to experiment.  Their model considers the dephasing of excitons and static energy disorder.  As previously described in Subsection~\ref{sec:qfi_dimer_spectroscopy}, the measured QFI is independent of the homogeneous line broadening in the optical response due to exciton-phonon interactions.  However, the effect of static disorder on the measured QFI has yet to be addressed which we incorporate in our analysis here.  

Static variations of the chemical environment in which the sites are embedded causes perturbations in both sites energies (diagonal disorder) and couplings (off-diagonal disorder).  Assuming such variations are uncorrelated among the sites, the electronic Hamiltonian can be modeled with site energies given by $E_{n}=E+\delta E$ where $\delta E$ is a Gaussian random variable $N\left(0,\sigma_{\delta E}^{2}\right)$.  The dipole coupling between sites is given by $J_{mn}=-J^{\prime}/\left|x_{m}-x_{n}\right|^{3}$ where $x_{m}$ is the position of site $m$ given by $x_{m}=ma + \delta x$ where $a$ is the lattice spacing that defines the mean distance between nearest-neighbors and $\delta x$ is a Gaussian random variable $N\left(0, \sigma_{\delta x}^{2}\right)$.  The QFI (Eq.~\ref{qfi_thermal_exciton}) is calculated for each realization and the final result is an average over all $M$ realizations.
\begin{figure}[h]
    \centering
    \includegraphics[width=0.45\textwidth]{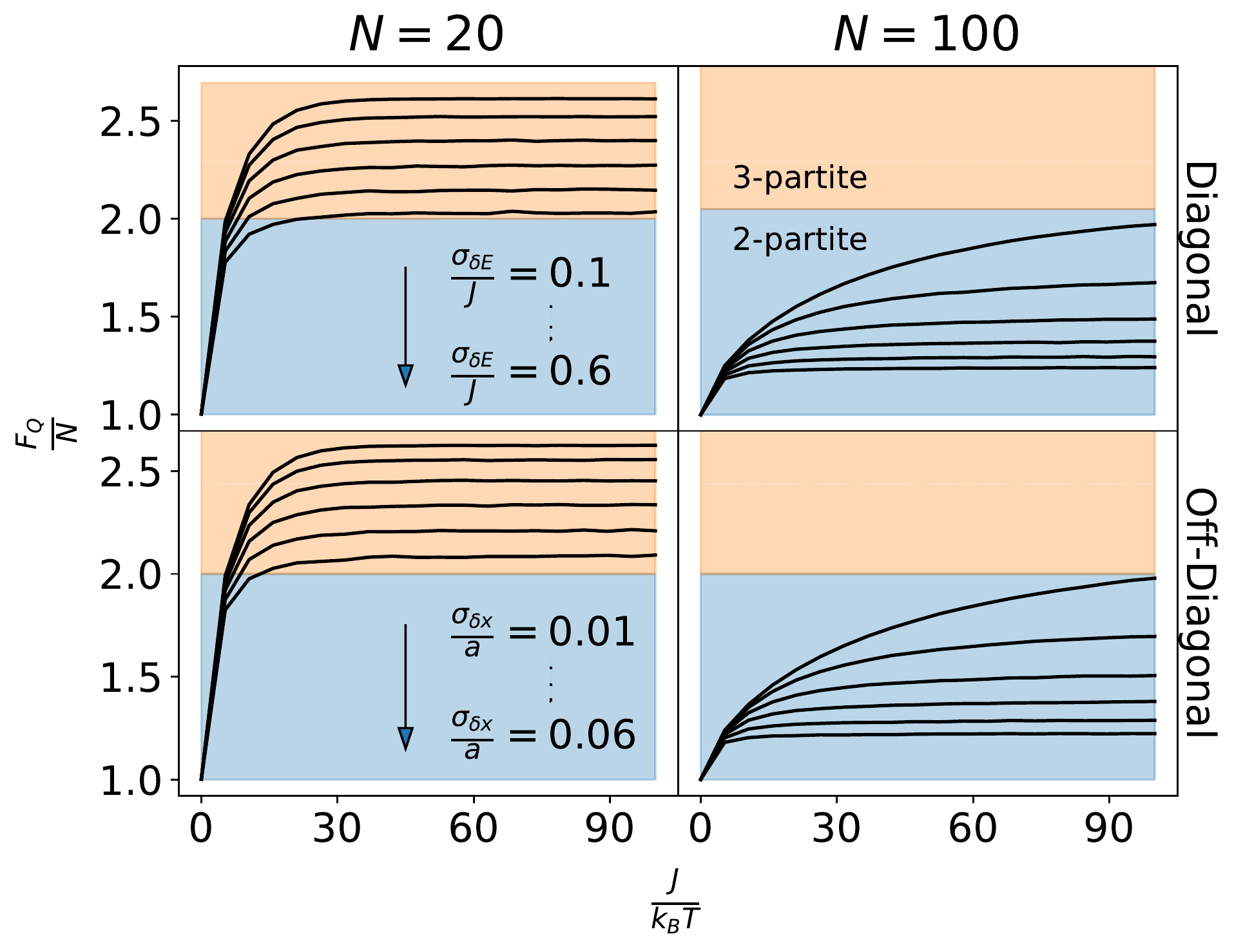}
    \caption{QFI per number of sites $F_{Q}/N$ for linear dye aggregates as a function of $J/k_{B}T$ (where $J=J^{\prime}/a^{3}$) for different degrees of disorder.  Diagonal (top panels) and off-diagonal (bottom panels) disorder are considered separately. $F_{Q}/N$ is calculated for six equally-spaced disorder parameters: $\sigma_{\delta E}/J=0.1$ to $0.6$ (diagonal) and $\sigma_{\delta x}/a=0.01$ to $0.06$ (off-diagonal).  Aggregate lengths are $N=20$ (left panels) and $N=100$ (right panels).  Results are averaged over $M=10^{4}$ realizations.  Disorder lessens the degree of entanglement, but even 3-partite entanglement is realizable in the presence of relatively strong disorder.} 
    \label{fig:qfi_linear_aggregate_disorder}
\end{figure}

In our predictions shown in Fig.~\ref{fig:qfi_linear_aggregate_disorder}, aggregate sizes of $N=20$ and $N=100$ are considered.  Diagonal and off-diagonal disorder are calculated separately in order to isolate the effects of each type.  The general trend of the QFI is that it increases as temperature decreases and then levels off at a maximum value for a given amount of disorder.  Disorder is found to lessen the degree of entanglement (\textit{i.e.}, Anderson localization).  More importantly, 3-partite entanglement is potentially realizable in the presence of strong disorder.  Fig.~\ref{fig:qfi_linear_aggregate_disorder} shows the witnessed QFI with site energies varying with $\sigma_{\delta E}=0.1J$ to $0.6J$ (top panels), or coupling between nearest-neighbors $\sigma_{J_{m,m\pm 1}}={\sim}0.04J$ to $0.3J$ and second nearest-neighbors $\sigma_{J_{m,m\pm 2}}={\sim}0.003J$ to $0.02J$ (bottom panels).  Reported disorder parameters for PIC aggregates are $\sigma_{\delta E}/J=0.128$ for PIC-Cl and $\sigma_{\delta E}/J=0.249$ for PIC-F.\cite{heijs2005decoherence}  Together with coupling $J=600$ cm$^{-1}$ and an aggregate size of $N=20$, 3-partite entanglement may be witnessed at a wide range of feasible temperatures, \textit{e.g.}, ${\sim}17$ K to 58 K (corresponding to $J/k_{B}T=50$ to 15).  Experiments that are capable of recovering trends in the predicted QFI, particularly in the regime of temperatures showing a dramatic change in the QFI and a crossover between 2- and 3-partite entanglement, would be particularly desirable for validating the present theory.

\subsection{Experimental Realization and Challenges}

An essential ingredient assumed in the present theory is that the stimulated emission and excited state absorption spectra of the probed state can be adequately discriminated in the overall response.  The challenge here is in avoiding an overlap of stimulated emission and ground state bleach, the latter being the response of aggregates in the ground state rather than the entangled excited state.  In small rigid systems, the 0-0 band of the ground state bleach may overlap with the 0-0 band in stimulated emission. But in many systems with more flexibility, the Stokes' shift is sufficiently large, ensuring that each signal can be spectrally resolved.  Excited state absorption appears in a lower energy regime and is thus not a limiting factor in this regard. 

In experiment, the quantity that is measured is the molar extinction coefficient $\epsilon[\frac{Lcm^{-1}}{mol}]$ determined by the Beer-Lambert Law $\frac{I}{I_{0}}=10^{-\epsilon Cl}$ where $I[\frac{Js^{-1}}{cm^{2}}]$ is the intensity of light, $C[\frac{mol}{L}]$ is the molar concentration, and $l[cm]$ is the path length of the probing field through the sample.  The differential change of intensity per unit length is related to the energy absorbed $\frac{\mathrm{d}I}{\mathrm{d}x} = -n\alpha I$ where $n[cm^{-3}]$ is the aggregate density and $\alpha[cm^{2}]$ is the absorption cross section.  These relations give $\epsilon=\frac{N_{A}\alpha}{\ln\left(10\right)10^{3}N}$ where $N_{A}$ is Avogadro's constant and $N$ is the number of molecules per aggregate.  The absorption cross section is given by $\alpha = \frac{\hbar\omega}{I_{0}} \sum_{m,n} p_{n}W_{mn}$ where $p_{n}$ is the population of $\ket{\epsilon_{n}}$ given by Eq.~\ref{eq:thermal_state}, $I_{0}=\frac{1}{2}\epsilon_{0}c E_{0}^{2}$ is the total intensity of the incident field, and $W_{mn}[s^{-1}]=\frac{\pi E_{0}^{2}}{2\hbar^{2}}|\langle \epsilon_{m}|\hat{E}\cdot \vec{\mu}|\epsilon_{n}\rangle|^{2}\delta\left(\omega - \omega_{mn}\right)$ is the rate of transition from $\ket{\epsilon_{m}}$ to $\ket{\epsilon_{n}}$ due to the interaction between the system's dipole moment and the probing field with frequency $\omega$.  These relations can be used to solve for the (unitless) QFI per number of sites:
\begin{align}\label{eq:qfi_per_nsites_related_to_experiment}
\begin{split}
\frac{F_{Q}}{N}&=\frac{4}{N}\sum_{n}^{N} p_{n}\left<\epsilon_{n}|\mathcal{O}^{2}|\epsilon_{n}\right>\\
&=\frac{3\ln\left(10\right) 10^{3}\epsilon_{0}c\hbar}{\pi N_{A} \mu_{i}^{2}} \int \frac{\epsilon\left(\omega\right)}{\omega} \mathrm{d}\omega
\end{split}
\end{align}
The integral in Eq.~\ref{eq:qfi_per_nsites_related_to_experiment} is taken over the stimulated emission and excited state absorption parts of the spectrum.  Here, $\mathcal{O}=\frac{1}{2}\sum_{i}^{N}\sigma_{i}^{x}$ and $\mu_{i}[Ccm]$ is the magnitude of the single-site transition dipole moment.  Eq.~\ref{eq:qfi_per_nsites_related_to_experiment} assumes an isotropic field with uniform polarization in all directions resulting in $\left<\cos^{2}\theta\right>=1/3$.  In theory, Eq.~\ref{eq:qfi_per_nsites_related_to_experiment} is a valid expression for deducing the $n$-partite entanglement (Eq.~\ref{multipartite_entanglement_condition}).  Although it should be stated that a more rigorous formulation of the extracted QFI that considers the combined pump-probe signal is warranted.  Ref.~\cite{heijs2007ultrafast} provides expressions of the incoherent and coherent parts of the signal in linear aggregates, the former of which contains the relevant information.

An important issue regarding practicality of Eq.~\ref{eq:qfi_per_nsites_related_to_experiment} is the possible ambiguity in the QFI of a state determined based on a response that is ensemble-averaged. The results of Fig.~\ref{fig:qfi_linear_aggregate_disorder} were obtained with a fixed number of sites in each aggregate $N$.  However, in a realistic sample this may be a crude approximation as not all aggregates are of identical size; instead, aggregate size may be best described by some distribution.  A generalization of this work to handle such uncertainties may prove useful. Perhaps an easier first-step for measurements in bulk would be to observe the arctan-like dependence of the integrated response as a function of inverse temperature (Fig.~\ref{fig:qfi_linear_aggregate_disorder}), without classifying $n$-partite entanglement.  Single-aggregate spectroscopy eliminates variation on $N$ and may be a more reliable approach for assigning multipartite entanglement.\cite{hernando2006effect}

\section{Conclusions and Future Outlook}

Nonclassical correlations are a potential resource in engineered and biological systems.  There has been substantial progress in the foundations of nonclassical correlations from an information-theoretic standpoint, but experimental realization of such correlations, including entanglement, is notoriously difficult to achieve.  Our work presented a theory based on the quantum Fisher information (QFI) for measuring entanglement of molecular excitons--delocalized electronic states in molecular aggregates--from spectroscopy.  We showed that the optical response of a thermally equilibrated exciton state in the first-excitation subspace is a valid QFI.  The QFI of the probed state is encoded in its stimulated emission and excited state absorption spectra.  Our analyses of the molecular dimer, $N$-site linear aggregate with nearest-neighbor coupling, and $N$-site circular aggregate showed that the response of a dominant dipole-susceptible state effectively determines the witnessed entanglement.  Applying the theory to realistic linear dye J aggregates, we predict that 2- and 3-partite entanglement are realizable with ultrafast pump-probe experiments, assuming that the appropriate signals can be adequately discriminated in the overall response.

While this work is a step toward measuring nonclassical correlations in molecular excitons, ultimately it is the evolution of nonclassical correlations that is most insightful, providing a lens to the rich exciton-phonon dynamics.  In order to achieve this goal with an outlook on spectroscopic techniques, the proposed measure must be both expressed in terms of response function(s) and generalized to arbitrary mixed states that are inevitable due to decoherence.  It would be interesting to apply the present analysis to light-matter states of cavity QED\cite{kockum2019ultrastrong} and entangled spins in molecular architectures.\cite{gaita2019molecular}


\begin{acknowledgments}
This material is based upon work supported by the U.S. Department of Energy, Office of Basic Energy Sciences under contract DE-SC0020437. G.D.S. and A.E.S. acknowledge support from the Gordon and Betty Moore Foundation through Grant GBMF7114.  F.F. acknowledges financial support from the European Union's H2020 Maria Sklodowska Curie actions, Grant agreement No. 799408.
\end{acknowledgments}

\appendix

\section{QFI of Molecular Dimer in Thermal Exciton State}
\label{qfi_thermal_exciton_dimer}

We evaluate the QFI (Eq.~\ref{qfi_mixed_state}) in a thermal exciton state given by Eq.~\ref{eq:thermal_state} for a molecular dimer with generator Eq.~\ref{linear_two_mode_interferometer}.  The terms include
\begin{subequations}
\begin{multline}\label{eq:epsilon0_epsilon1}
4p_{1}\left|\left<\epsilon_{0}|\mathcal{O}|\epsilon_{1}\right>\right|^{2}=\frac{e^{\frac{1}{k_{B}T}\left|\frac{V}{\sin 2\theta}\right|}}{\mathcal{Z}}\left[1+2\left(n^{x}_{1}n^{x}_{2}+n^{y}_{1}n^{y}_{2}\right)\right.\\
\left.\times\sin\theta\cos\theta-\left\{\left(n^{z}_{1}\right)^{2}\sin^{2}\theta+\left(n^{z}_{2}\right)^{2}\cos^{2}\theta\right\}\right]
\end{multline}
\begin{multline}\label{eq:epsilon0_epsilon2}
4p_{2}\left|\left<\epsilon_{0}|\mathcal{O}|\epsilon_{2}\right>\right|^{2}=\frac{e^{-\frac{1}{k_{B}T}\left|\frac{V}{\sin 2\theta}\right|}}{\mathcal{Z}}\left[1-2\left(n^{x}_{1}n^{x}_{2}+n^{y}_{1}n^{y}_{2}\right)\right.\\
\times\left.\sin\theta\cos\theta-\left\{\left(n^{z}_{1}\right)^{2}\cos^{2}\theta+\left(n^{z}_{2}\right)^{2}\sin^{2}\theta\right\}\right]
\end{multline}
\end{subequations}
The terms $4p_{1}\left|\left<\epsilon_{1}|\mathcal{O}|\epsilon_{3}\right>\right|^{2}$  and $4p_{2}\left|\left<\epsilon_{2}|\mathcal{O}|\epsilon_{3}\right>\right|^{2}$ are equivalent to Eqs.~\ref{eq:epsilon0_epsilon1} and \ref{eq:epsilon0_epsilon2}, respectively, with $\sin\theta\leftrightarrow\cos\theta$.  Lastly,
\begin{multline}
\frac{4\left(p_{1}-p_{2}\right)^{2}}{p_{1}+p_{2}}\left|\left<\epsilon_{1}|\mathcal{O}|\epsilon_{2}\right>\right|^{2}=4\tanh^{2}\left(\frac{1}{k_{B}T}\left|\frac{V}{\sin 2\theta}\right|\right)\\
\times\sin^{2}\theta\cos^{2}\theta\left[\left(n^{z}_{1}\right)^{2}-\left(n^{z}_{2}\right)^{2}\right]
\end{multline}
Combining all terms,
\begin{multline}\label{eq:qfi_thermal_exciton_in_terms_of_bloch}
F_{Q}=2+2f\left(\theta\right)\left(n_{1}^{x}n_{2}^{x}+n_{1}^{y}n_{2}^{y}\right)+f^{2}\left(\theta\right)\left(n_{1}^{z}-n_{2}^{z}\right)^{2}\\
-\left[\left(n^{z}_{1}\right)^{2}+\left(n^{z}_{2}\right)^{2}\right]
\end{multline}
where $f\left(\theta\right)=\tanh\left(\frac{1}{k_{B}T}\left|\frac{V}{\sin 2\theta}\right|\right)\sin 2\theta$.  The $\hat{x}$ and $\hat{y}$ components must be parallel or antiparallel depending on the sign of $f\left(\theta\right)$ to maximize Eq.~\ref{eq:qfi_thermal_exciton_in_terms_of_bloch}.  Eq.~\ref{eq:max_qfi_thermal_exciton_dimer} is attained by setting $n_{1}^{x}n_{2}^{x}+n_{1}^{y}n_{2}^{y}=\text{sgn}\left[f\left(\theta\right)\right]-n_{1}^{z}n_{2}^{z}$ in Eq.~\ref{eq:qfi_thermal_exciton_in_terms_of_bloch} and maximizing with respect to $n_{1}^{z}$ and $n_{2}^{z}$, resulting in $n_{1}^{z}=n_{2}^{z}=0$.

\section{QFI of Linear Aggregate from Spectroscopy}
\label{qfi_linear_aggregate_derivation}
We evaluate the QFI of the states $\ket{\psi_{k}^{L}}$ (Eq.~\ref{eigenstates_linear_aggregate_nn_coupling}) with generator $\mathcal{O}=\frac{1}{2}\sum_{i}^{N}\sigma_{i}^{x}$.  Expressing $\sigma_{i}^{x}$ in terms of raising and lowering operators $\sigma_{i}^{x}=\sigma_{i}^{+}+\sigma_{i}^{-}$, $\left<\mathcal{O}\right>=0$ since $\left<m|\sigma^{+}_{i}|n\right>=\left<m|\sigma^{-}_{i}|n\right>=0$.  The QFI becomes $F_{Q}=4\left<\mathcal{O}^{2}\right>$.  An expansion of $\mathcal{O}^{2}$ gives
\begin{equation}\label{dipole_squared_generator}
\mathcal{O}^{2}=\frac{1}{4}\sum_{i,j}^{N}\sigma^{+}_{i}\sigma^{+}_{j}+\sigma^{+}_{i}\sigma^{-}_{j}+\sigma^{-}_{i}\sigma^{+}_{j}+\sigma^{-}_{i}\sigma^{-}_{j}
\end{equation}
The middle two terms preserve the number of excitations whereas the first and last terms connect states that differ by two excitations.  Only the middle two terms have non-zero expectation values.  We begin by evaluating these terms in the site basis: $\left<m|\sigma^{+}_{i}\sigma^{-}_{j}|n\right>$ and $\left<m|\sigma^{-}_{i}\sigma^{+}_{j}|n\right>$.  The first term is straightforward $\left<m|\sigma^{+}_{i}\sigma^{-}_{j}|n\right>=\delta_{im}\delta_{jn}$, resulting in
\begin{equation}\label{raise_lower_term}
\left<\sigma^{+}_{i}\sigma^{-}_{j}\right>=\frac{2}{N+1}\left[\sum_{n}\sin\left(\frac{\pi k n}{N+1}\right)\right]^{2}
\end{equation}
whereas for the second term we must consider combinations $i$, $j$.  For $i=j$, $\left<m|\sigma^{-}_{i}\sigma^{+}_{j}|n\right> = \left<m|\left[\sigma^{-}_{i}, \sigma^{+}_{j}\right]|n\right> + \left<m|\sigma^{+}_{j}\sigma^{-}_{i}|n\right> = \delta_{ij}\delta_{mn}\left(1-2\delta_{jm}\right)+\delta_{ij}\delta_{jm}\delta_{mn}$.  Here, we have used $\left[\sigma_{i}^{+},\sigma_{i}^{-}\right]=\sigma_{i}^{z}$ and $\left<m|\sigma^{z}_{i}|n\right>=\delta_{mn}\left(2\delta_{im}-1\right)$.  For $i\ne j$, $\left<m|\sigma^{-}_{i}\sigma^{+}_{j}|n\right>=\left<m|\sigma^{+}_{j}\sigma^{-}_{i}|n\right>=\delta_{jm}\delta_{in}-\delta_{ij}\delta_{jm}\delta_{mn}$.  Combining these parts,
\begin{multline}\label{lower_raise_term}
\left<\sigma^{-}_{i}\sigma^{+}_{j}\right>=\frac{2\left(N-2\right)}{\left(N+1\right)}\sum_{n}^{N}\sin^{2}\left(\frac{\pi k n}{N+1}\right)\\
+\frac{2}{N+1}\left[\sum_{n}^{N}\sin\left(\frac{\pi k n}{N+1}\right)\right]^{2}
\end{multline}
Eq.~\ref{eq:qfi_linear_aggregate_nn_coupling} is recovered by summing Eqs.~\ref{raise_lower_term} and \ref{lower_raise_term} and evaluating the summations
\begin{subequations}
\begin{equation}
\left[\sum_{n}^{N} \sin\left(\frac{\pi k n}{N+1}\right)\right]^{2}=\frac{1-\left(-1\right)^{k}}{2}\frac{\sin^{2}\left(\frac{\pi k N}{2\left(N+1\right)}\right)}{\sin^{2}\left(\frac{\pi k}{2\left(N+1\right)}\right)}
\end{equation}
\begin{equation}
\sum_{n}^{N}\sin^{2}\left(\frac{\pi k n}{N+1}\right)=\frac{N+1}{2}
\end{equation}
\end{subequations}

\section{QFI of Circular Aggregate from Spectroscopy}
We evaluate $\mathcal{O}=\frac{1}{2}\sum_{i}^{N}\sigma_{i}^{x}$ in the states $\ket{\psi_{k}^{C}}$ (Eq.~\ref{eq:eigenstates_circular_aggregate}).  Similar to Appendix~\ref{qfi_linear_aggregate_derivation}, $F_{Q}=4\left<\mathcal{O}^{2}\right>$ with $\mathcal{O}$ given by Eq.~\ref{dipole_squared_generator}. Evaluating the two non-zero terms $\left<\sigma^{+}_{i}\sigma^{-}_{j}\right>$ and $\left<\sigma^{-}_{i}\sigma^{+}_{j}\right>$ gives
\begin{equation}\label{qfi_circular_aggregate_not_reduced}
F_{Q}^{k}=\left(N-2\right)+\frac{2}{N}\frac{\sin^{2}\left(\pi k\right)}{\sin^{2}\left(\pi k/ N\right)}
\end{equation}
where we have used the summation
\begin{equation}
\sum_{n}\sum_{m}e^{2\pi i k\left(n-m\right)/N}=\frac{\sin^{2}\left(\pi k\right)}{\sin^{2}\left(\pi k/ N\right)}
\end{equation}
Eq.~\ref{qfi_circular_aggregate_not_reduced} is equivalent to
\begin{equation}\label{qfi_circular_aggregate_reduced}
F_{Q}^{k}=
  \begin{cases}
      3N-2 & \text{for $k=0$}\\
      N-2 & \text{otherwise}
    \end{cases}  
\end{equation}

\section{Maximum  QFI of $k=0$ State in Circular Aggregate}
\label{maximum_qfi_k=0_state_circular_aggregate}
We evaluate Eq.~\ref{linear_two_mode_interferometer} in the state $\ket{\psi_{0}^{C}}=\frac{1}{\sqrt{N}}\sum_{n}^{N} \ket{n}$ and show that the maximum QFI $F_{Q}^\text{max}=3N-2$ can be obtained with $\mathcal{O}=\frac{1}{2}\sum_{i}^N \sigma_{i}^{x}$.  Expanding $\left<\mathcal{O}^{2}\right>$, there are nine terms: $\sigma^{x}_{i}\sigma^{x}_{j}$, $\sigma^{x}_{i}\sigma^{y}_{j}$, $\sigma^{x}_{i}\sigma^{z}_{j}$, etc.  The expectation values of $\sigma^{x}_{i}\sigma^{z}_{j}$, $\sigma^{z}_{i}\sigma^{x}_{j}$, $\sigma^{y}_{i}\sigma^{z}_{j}$, and $\sigma^{z}_{i}\sigma^{y}_{j}$ are zero because they connect states with a different number of excitations.

The terms $\left<\sigma^{x}_{i}\sigma^{y}_{j}\right>$ and $\left<\sigma^{y}_{i}\sigma^{x}_{j}\right>$ combine to give 
\begin{equation}\label{alpha}
\alpha=i\left(n^{x}_{i}n^{y}_{j}-n^{y}_{i}n^{x}_{j}\right)\left[\left<\sigma^{+}_{i}\sigma^{-}_{j}\right>-\left<\sigma^{-}_{i}\sigma^{+}_{j}\right>\right]
\end{equation}
Terms in $\left<\sigma^{x}_{i}\sigma^{y}_{j}\right>$ and $\left<\sigma^{y}_{i}\sigma^{x}_{j}\right>$ that change the number of excitations are not shown in Eq.~\ref{alpha} since their expectation values are zero.  Eq.~\ref{alpha} is zero since for $i=j$, the term $n^{x}_{i}n^{y}_{i}-n^{y}_{i}n^{x}_{i}=0$, whereas for $i\ne j$, $\left<\sigma^{+}_{i}\sigma^{-}_{j}\right>^{*}=\left<\sigma^{-}_{i}\sigma^{+}_{j}\right>$.  The term in $\left[\cdots\right]$ becomes $2i\text{Im}\left[\left<\sigma^{+}_{i}\sigma^{-}_{j}\right>\right]=0$.

The terms $\left<\sigma^{x}_{i}\sigma^{x}_{j}\right>$ and $\left<\sigma^{y}_{i}\sigma^{y}_{j}\right>$ combine to give 
\begin{equation}\label{beta}
\beta=\left(n^{x}_{i}n^{x}_{j}+n^{y}_{i}n^{y}_{j}\right)\left[\left<\sigma^{+}_{i}\sigma^{-}_{j}\right>+\left<\sigma^{-}_{i}\sigma^{+}_{j}\right>\right]  
\end{equation}
Both $\left<m|\sigma^{+}_{i}\sigma^{-}_{j}|n\right>$ and $\left<m|\sigma^{-}_{i}\sigma^{+}_{j}|n\right>$ have been evaluated in Appendix~\ref{qfi_linear_aggregate_derivation}.  Eq.~\ref{beta} becomes
\begin{equation}\label{xx_yy_term}
\beta=\frac{3}{N}\sum_{i,j}^{N} n^{x}_{i}n^{x}_{j}+n^{y}_{i}n^{y}_{j}-\frac{2}{N}\sum_{i}^{N}n^{x}_{i}n^{x}_{i}+n^{y}_{i}n^{y}_{i}
\end{equation}

The relation $\left[\sigma_{i}^{+},\sigma_{i}^{-}\right]=\sigma_{i}^{z}$ is used to obtain $\left<m|\sigma^{z}_{i}\sigma^{z}_{j}|n\right>=\delta_{mn}\delta_{in}\delta_{jn}-\delta_{mn}\delta_{in}\left(1-\delta_{jn}\right)-\delta_{mn}\delta_{jn}\left(1-\delta_{in}\right)+\delta_{mn}\left(1-\delta_{in}\right)\left(1-\delta_{jn}\right)$.  Reducing further, $\left<m|\sigma^{z}_{i}\sigma^{z}_{j}|n\right>=4\delta_{mn}\delta_{in}\delta_{jn}-4\delta_{mn}\delta_{in}+\delta_{mn}$.  The term $\gamma=n^{z}_{i}n^{z}_{j}\left<\sigma^{z}_{i}\sigma^{z}_{j}\right>$ becomes
\begin{equation}\label{zz term}
\gamma=\frac{4}{N}\sum_{i}^{N}n^{z}_{i}n^{z}_{i}-\frac{4}{N}\sum_{i,j}^{N}n^{z}_{i}n^{z}_{j}+\sum_{i,j}^{N}n^{z}_{i}n^{z}_{j}
\end{equation}

Summing Eq.~\ref{xx_yy_term} and Eq.~\ref{zz term} completes the calculation of $4\left<\mathcal{O}^{2}\right>$.  The term $\left<\mathcal{O}\right>^{2}$ has a single non-zero term from $\left<\sigma^{z}_{i}\right>$
\begin{equation}\label{expectation_squared}
4\left<\mathcal{O}\right>^{2}=\left(\frac{2-N}{N}\right)^{2}\sum_{i,j}n^{z}_{i}n^{z}_{j}
\end{equation}

Combining Eqs.~\ref{xx_yy_term}, \ref{zz term}, and \ref{expectation_squared}, the QFI becomes
\begin{equation}\label{general_qfi_circular_aggregate_not_reduced}
F_{Q}=3N-2+\frac{6}{N}\sum_{i}^{N}n^{z}_{i}n^{z}_{i}-\frac{3N+4}{N^{2}}\sum_{i,j}^{N}n^{z}_{i}n^{z}_{j}
\end{equation}
The $\hat{x}$ and $\hat{y}$ components of all $\vec{n}_{i}$ are set parallel to one another to maximize Eq.~\ref{general_qfi_circular_aggregate_not_reduced}, $n^{x}_{i}n^{x}_{j}+n^{y}_{i}n^{y}_{j}=1-n_{i}^{z}n_{j}^{z}$.  Maximizing Eq.~\ref{general_qfi_circular_aggregate_not_reduced} with respect to $n_{i}^{z}$ gives the system of equations
\begin{equation}\label{eq:system_of_equations}
\frac{3N-4}{N^{2}}n_{i}^{z}-\frac{3N+4}{N^{2}}\sum_{j\ne i}^{N}n_{j}^{z}=0
\end{equation}
which can be recast in terms of a circulant coefficient matrix $C$
  \begin{equation}\label{eq:circulant_matrix}
    \begin{pmatrix}
      c_{1} & c_{N} & \cdots & c_{2} \\
      c_{N} & c_{1} & \cdots & c_{3} \\
      \vdots & \vdots & \ddots & \vdots \\
      c_{2} & c_{3} & \cdots & c_{1}
    \end{pmatrix}
    \begin{pmatrix}
      n_{1}^{z} \\
      n_{2}^{z} \\
      \vdots    \\
      n_{N}^{z}
    \end{pmatrix}
    =0
  \end{equation}
with matrix elements $c_{1}=\left(3N-4\right)/N^{2}$ and $c_{i\ne 1}=-\left(3N+4\right)/N^{2}$.  For Eq.~\ref{eq:circulant_matrix} to host nontrivial solutions, the determinant of $C$
\begin{equation}
\text{det}\left(C\right)=\prod_{j=0}^{N-1}\left(c_{1}+c_{N}\omega^{j}+c_{N-1}\omega^{2j}+\cdots +c_{2}\omega^{\left(N-1\right)j}\right)
\end{equation}
where $\omega=\exp\left(2\pi i/N\right)$ must be zero.  Solving further, 
\begin{align}\label{eq:determinant}
\begin{split}
\text{det}\left(C\right)&=\left(\frac{3N-4}{N^{2}}\right)^{N}+\left(-\frac{3N+4}{N^{2}}\right)^{N}\sum_{j=1}^{N-1}\omega^{j\sum_{k=0}^{N-1}k}\\
&=\left(\frac{3N-4}{N^{2}}\right)^{N}+\left(-\frac{3N+4}{N^{2}}\right)^{N}\sum_{j=1}^{N-1}r^{j}
\end{split}
\end{align}
where $r=\exp\left[i\pi\left(N-2\right)\right]=\left(-1\right)^{N}$.  The determinant reduces to
\begin{equation}
\text{det}\left(C\right)=
  \begin{cases}
      \left(\frac{3N-4}{N^{2}}\right)^{N}+\left(\frac{3N+4}{N^{2}}\right)^{N}\left(N-1\right) & \text{$N$ even}\\
      \left(\frac{3N-4}{N^{2}}\right)^{N} & \text{$N$ odd}
    \end{cases}  
\end{equation}
Since $\text{det}\left(C\right)\ne 0$, the only solution of Eq.~\ref{eq:system_of_equations} is the trivial one with $n_{i}^{z}=0$.   The maximum QFI is $F_{Q}^\text{max}=3N-2$.  All generators that satisfy $\vec{n}_{i}\cdot\vec{n}_{j}=1$ and $n^{z}_{i}=0$ will give the maximum QFI, among which is $\mathcal{O}=\frac{1}{2}\sum_{i}^N \sigma^{x}_{i}$.

\bibliography{references}

\end{document}